\documentclass[11pt]{article}

\usepackage{microtype} % microtypography
\usepackage{booktabs}  % tables
\usepackage{url}  % urls
\usepackage{csquotes}
\usepackage{listings}

\usepackage[table,xcdraw]{xcolor}
\usepackage{amsmath}
\usepackage{amsthm}
\usepackage{xcolor}
\definecolor{codegreen}{rgb}{0,0.6,0}
\definecolor{backcolor}{rgb}{0.95, 0.95, 0.96}
\definecolor{codepurple}{rgb}{0.58,0,0.82}
\definecolor{codegray}{rgb}{0.5,0.5,0.5}

\lstdefinestyle{mystyle}{
backgroundcolor=\color{backcolor},
keywordstyle=\color{codegreen},
basicstyle=\ttfamily\footnotesize,
upquote=true,
stringstyle=\color{codepurple},
numberstyle=\color{codegray},
numbers=left
}
\usepackage[preprint]{automl}
\usepackage{graphicx}
\usepackage{natbib}

\newcommand{\subf}[2]{%
  {\small\begin{tabular}[t]{@{}c@{}}
  #1\\#2
  \end{tabular}}%
}

\title{MindOpt Tuner: Boost the Performance of Numerical Software by Automatic Parameter Tuning}

\author[1]{\nameemail{Mengyuan Zhang}{marvin.zmy@alibaba-inc.com}}
\author[1]{\nameemail{Wotao Yin}{wotao.yin@alibaba-inc.com}}
\author[1]{\nameemail{Mengchang Wang}{mengchang.wmc@alibaba-inc.com}}
\author[1]{\nameemail{Yangbin Shen}{yangbin.syb@alibaba-inc.com}}
\author[1]{\nameemail{Peng Xiang}{xiangpeng.xp@alibaba-inc.com}}
\author[1]{\nameemail{You Wu}{wy186633@alibaba-inc.com}}
\author[1]{\nameemail{Liang Zhao}{yuanyue.zl@alibaba-inc.com}}
\author[1]{\nameemail{Junqiu Pan}{panjunqiu.pjq@alibaba-inc.com}}
\author[1]{\nameemail{Hu Jiang}{qinweng.jh@taobao.com}}
\author[1]{\nameemail{KuoLing Huang}{kuoling.huang@alibaba-inc.com}}
\affil[1]{Decision Intelligence Lab, Alibaba Group}
\hypersetup{%
  pdfauthor={AutoML}, % will be reset to "Anonymous" unless the "final" package option is given
  pdftitle={Documentation for the MindOpt Tuner},
  pdfsubject={Documentation for MindOpt Tuner},
  pdfkeywords={HPO, LaTeX, style}
}

\begin{document}

%% to show revisions:
% \setreviewson

%% to turn off revisions:
%\setreviewsoff

\maketitle

\begin{abstract}
  Numerical software is usually shipped with built-in hyperparameters. By carefully tuning those hyperparameters, significant performance enhancements can be achieved for specific applications. We developed MindOpt Tuner, a new automatic tuning tool that supports a wide range of numerical software, including optimization and other solvers. MindOpt Tuner uses elastic cloud resources, features a web-based task management panel and integration with ipython notebook with both command-line tools and Python APIs. Our experiments with COIN-OR Cbc, an open-source mixed-integer optimization solver, demonstrate remarkable improvements with the tuned parameters compared to the default ones on the MIPLIB2017 test set, resulting in over 100x acceleration on several problem instances. Additionally, the results demonstrate that Tuner has a higher tuning efficiency compared to the state-of-the-art automatic tuning tool SMAC3. 
\end{abstract}

\section{Introduction}
Numerical software is designed for solving complex problems that involve numerical calculations. These computations typically involve algorithms with many configurable hyperparameters. For instance, Cbc by COIN-OR \cite{cbc}, one of the state-of-the-art open-source numerical software for solving mixed integer linear programming (MILP), has approximately 160 configurable hyperparameters that control the behavior of algorithms inside. Depending on the numerical structures of the MILP instances, different hyperparameter combinations (referred to as hyperparameters throughout this paper) are preferred to achieve optimal performances. Finding the best hyperparameters for a class of MILP instances may be one of the key factors in the successful application of an optimization solver.

Due to the lack of prior knowledge regarding the relationship between hyperparameters and actual performance on specific instances, we may have to attempt to solve typical instances multiple times with different hyperparameters. This process can be tedious and time-consuming, even with when using the built-in tuning tools designed to search for better hyperparameters.

We have developed a new tuning tool, MindOpt Tuner, which can automatically evaluate large amounts of hyperparameters utilizing cloud resources and accelerate the tuning progress with massive parallelism. 
Currently, MindOpt Tuner V0.9 has provided full support for optimization solvers, a typical class of numerical software, 
% (a typical group of supports tuning for various open-source/commercial solvers such as Cbc, IBM ILOG CPLEX\footnote{https://www.ibm.com/products/ilog-cplex-optimization-studio}, and MindOpt solvers\footnote{https://opt.alibabacloud.com/\#/portal} 
to enhance their performances on optimization problems, including reducing solving time and/or the optimality gap of the solution. 
% Moreover, MindOpt Tuner \textcolor{blue}{supports the tuning against problems modeled with MindOpt APL}, a modeling language in the MindOpt Suite\footnote{https://opt.alibabacloud.com/\#/portal}. 
In addition to optimization solvers, MindOpt Tuner supports tuning any 3rd-party numerical software. % through proper configurations. 
The homepage of MindOpt Tuner is located at:

\begin{center}
  \url{https://opt.alibabacloud.com/#/tuner/quicktry}
\end{center}

In this paper:
\begin{enumerate}
    \item We introduce MindOpt Tuner, an automatic hyperparameter tuning tool for numerical software that leverages the elasticity of the cloud computing resources.
    \item We present three approaches for using MindOpt Tuner, which enable diverse groups of users to execute and manage their tuning tasks in a flexible and convenient manner.
    \item We provide empirical evaluations of MindOpt Tuner, which validate its effectiveness and cost-efficiency in enhancing the performance of the optimization solver. 
\end{enumerate}

This paper is organized as follows. We first discuss related works on hyperparameter tuning algorithms and tools in Section \ref{sec:realted_works}. In Section \ref{sec:mindopt_tuner}, we introduce the architecture and algorithm of MindOpt Tuner, and present three approaches for using MindOpt Tuner, i.e., web-based task management, command-line tool, and Python APIs. In Section \ref{sec:numerical}, we provide empirical evaluations for MindOpt Tuner by tuning Cbc's parameters on the MIPLIB2017 Benchmark set \cite{miplib2017} and compare its performance to SMAC3 \cite{smac2022}. 
% Further discussions are provided in Section \ref{sec:discussion}. 
Finally, we conclude the paper in Section \ref{sec:conclusion}.
For brevity, we will hereafter use the terms ``software'' and ``algorithm'' in place of "numerical software" and "tuning algorithm", respectively.

\section{Related Work}\label{sec:realted_works}
In this section, we provide a non-comprehensive review of the state-of-the-art tuning methods and tools. Please refer to \cite{hutter2019automated, YANG2020295} for a comprehensive review of hyperparameter tuning techniques and tools.
Basic methods for hyperparameter tuning include random search \cite{random_search} and grid search \cite{montgomery2017design}: the former samples a number of random samples independently, usually based on a specified probability distribution; the latter partitions the search space into a grid and evaluates the candidate hyperparameters corresponding to each grid point in turn, with sampling complexity increasing exponentially with the dimensionality of the hyperparameters. More advanced algorithms include evolution strategy based ones that maintain a population of candidate hyperparameters that are randomly mutated and selected for survival based on their fitness values \cite{cma2003}, and Bayesian optimization methods that use Bayesian inference to construct a probabilistic model (e.g., Gaussian process, random forest) of the mapping from hyperparameters to numerical performance, which guides the search for optimal hyperparameters \cite{boreview}. Some other algorithms include gradient estimation based method \cite{zoopt2017} and tree-search based method \cite{lamcts2020}, etc.

Based on the various kinds of algorithms, several hyperparameter tuning tools and products have been proposed. 
ParamILS \cite{ParamILS} is a local search-based method that iteratively revises a candidate set of hyperparameters to get the optimal one.
The \verb|irace| \cite{irace2016} combines random sampling and statistical tests to quickly identify promising hyperparameters and eliminate poor ones.
MOSAIC \cite{mosaic2019} applies the Monte-Carlo Tree Search algorithm over the search space of hyperparameters, with its performance largely influenced by the parameters of MOSAIC itself. Bayesian Optimization has been widely used by the algorithm in a variety of popular hyperparameter tuning tools, including SMAC3 \cite{smac2022}, BOHB \cite{bohb2018}, HEBO \cite{hebo2022}, SigOpt \cite{sigopt}, HyperOpt \cite{hyperopt2013} and Dragonfly \cite{dragonfly2020}, etc. 

Some more advanced hyperparameter tuning tools have multiple algorithms included and are integrated with techniques such as early stopping and parallelization of hyperparameter evaluations \cite{raytune2018,optuna2019,vizier2017,nevergrad2018}. Ray Tune \cite{raytune2018}, for instance, is an open-source hyperparameter tuning library (mainly targeted at machine learning models) that synthesizes a bunch of algorithms. It is built on top of Ray, a popular open-source framework for building distributed applications, and can be scaled to large clusters and provide efficient parallelization of the tuning job. For most commercial hyperparameter tuning tools, e.g., SigOpt, the algorithm is run remotely. The evaluations of hyperparameters can be conducted at the user side and the evaluation results are submitted to the remote server via relevant APIs. Some AI platforms such as Google's Vertex AI, Amazon's SageMaker, and Microsoft Azure's Machine Learning Platform also provide tuning services through which users can run both the algorithm and the hyperparameter evaluations using cloud resources.

The majority of existing hyperparameter tuning tools are dedicated to use in machine learning applications, for instance, to improve the accuracy of machine learning models. However, there are some key differences between tuning for machine learning models and tuning for software. For example, processing different data samples during model training normally has equal costs in machine learning applications, and can be easily done in a parallel manner. However, when tuning for an optimization solver, each problem instance needs to be solved under different settings of hyperparameters. The time cost may vary a lot for different instances and for different hyperparameters, which brings greater challenges for improving the tuning performance under the guarantee of tuning efficiency.
Moreover, the dimensionality of hyperparameters for machine learning models is relatively small (usually less than 20), which is far less than the number of tunable hyperparameters for an optimization solver, which ranges from 100+ to 1000+ depending on the solvers.
Among the tools mentioned above, only SMAC3, ParamILS, and \verb|irace| are equipped with well-designed features for tuning numerical algorithms or software. They have devised a mechanism to improve the efficiency when the tuning objective is to search for hyperparameters that perform well across a large set of instances in general. However, these three tools are not designed to be directly used with elastic cloud resources.
Some software may provide their own tuners. For example, commercial optimization solvers such as CPLEX \cite{cplex2009v12}
% \footnote{https://www.ibm.com/products/ilog-cplex-optimization-studio}
, Gurobi \cite{gurobi}
% \footnote{https://www.gurobi.com/}
, and Copt \cite{copt}
% \footnote{https://www.shanshu.ai/copt}
have their own built-in tuners. Nonetheless, they are limited to tuning for their own solvers and can only be executed locally.

To the best of our knowledge, the MindOpt Tuner proposed in this paper is the first cloud-based tuning tool that provides tuning services to general numerical software, including open-source or commercial optimization solvers. 
% In particular, MindOpt Tuner can be used in conjunction with MindOpt APL modeling language, enabling parameter tuning for all the integrated solvers. 
It leverages the elasticity of cloud computing resources to reduce the resource cost of tuning tasks and applies a hierarchical surrogate modeling method to enhance the efficiency of the tuning algorithm.
% As a part of the MindOpt Suite, MindOpt Tuner is designed with several features that offer more convenience to users of optimization solvers. 
A well-developed command-line tool enables users to easily specify the model data and task configurations, such as the time limit of tuning process and the logging level. Additionally, it features a built-in log parser module that can extract abundant information from different fields of the software's output logs (customized by the users), which will be utilized by the algorithm to boost the software performance. 
% Moreover, MindOpt Tuner will support to be used in conjunction with MindOpt APL modeling language, enabling parameter tuning for all the integrated solvers.

\section{MindOpt Tuner}\label{sec:mindopt_tuner}
\subsection{Overview}
\begin{figure}[!ht]
  \centering
  \includegraphics[clip, trim=3cm 3.5cm 2cm 3cm,width=0.9\textwidth]{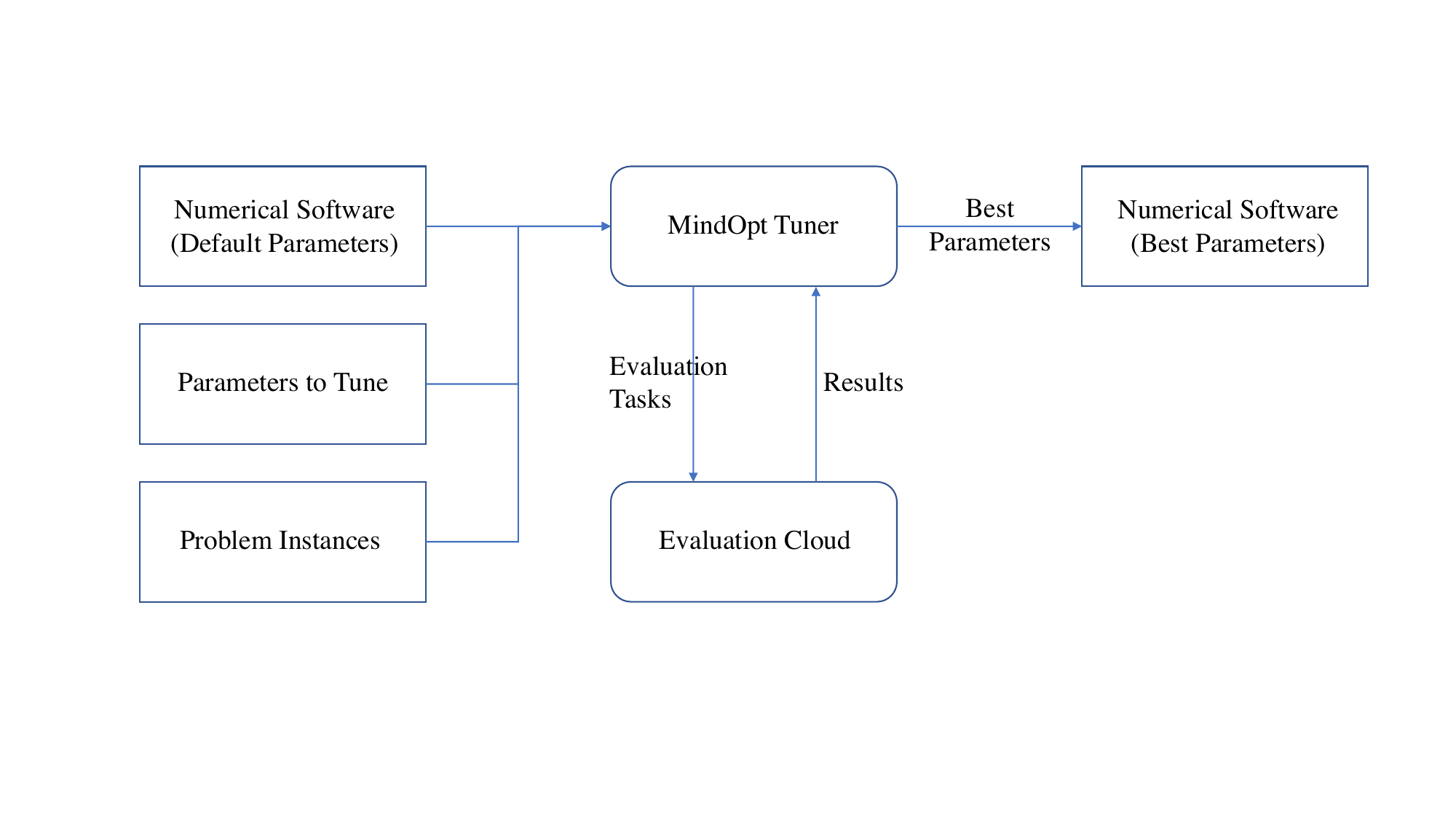}
  \caption{MindOpt Tuner Overview}
  \label{fig:workflow}
\end{figure}

The workflow of MindOpt Tuner is illustrated in Figure \ref{fig:workflow}. To create a tuning task, users need to first specify the target software, a set of hyperparameters that control the behavior of the software, and the data that describe the problem(s) for which the software will be tuned \footnote{Optimization solvers typically support problem instances given in the formats of \texttt{.mps}, \texttt{.lp}, and \texttt{.nl}.}. 
During the execution of the tuning task, MindOpt Tuner will generate and evaluate different samples of hyperparameters iteratively: 
The algorithm will solve the problem(s) using the current hyperparameters and observe the performance. Based on the collect performance information of all evaluated hyperparameters, the algorithm will then generate and evaluate new sample(s) of hyperparameters trying to improve the current best value of the tuning objective. The process is repeated until certain stopping conditions are reached..

Upon completion of the tuning task, the recommended (nearly optimal) hyperparameters of the software will be provided to the users. A collection of result files will also be available, offering comprehensive information on the progress and outcome of the task. For instance, the evaluation history of hyperparameters can be further utilized for result visualization and analysis. 

\subsection{Cloud Architecture}
MindOpt Tuner is connected to a cloud for parameter evaluations. In addition, MindOpt Tuner has a Web interface, a command-line interface on an online terminal, as well as a Python interface. The cloud brings multiple advantages:

\paragraph{Scalability} The computational resources required for hyperparameter tuning can vary significantly depending on the complexity and size of the numerical problems and the dimensionality of hyperparameters to tune. Cloud platforms allow us to easily scale up (or down) resources to meet these demands. In other words, we can take advantage of more computing power when determining large-scale hyperparameters over difficult problem instances and save costs when we don't. Some tuning projects can last weeks, and Alibaba Cloud comes with automated backups and recovery. 

\paragraph{Parallelization} We use model-based global optimization algorithms, which allow us to run multiple evaluation tasks with different hyperparameters concurrently. Cloud-based environments provide access to multiple workers simultaneously, making it easier to run these instances in parallel and thus helping identify the optimal hyperparameters more quickly.

\paragraph{Cost-effectiveness} Some users need tuning services only when they are unhappy with their current solution times. Once they are satisfied, they will no longer need to run tuning tasks. On-premise deployment of a system for tuning may require substantial upfront investments in hardware and software, ongoing maintenance costs, and possibly excess capacity. Cloud-based solutions generally follow a pay-as-you-go model, which can be more cost-effective.

\paragraph{Easy access} Since it is cloud-based, MindOpt Tuner is accessible from anywhere and not tied to a specific machine or local network. It provides several interfaces through Alibaba Cloud, including web, command line, and Python. This allows for more flexibility for the user. In addition, MindOpt Tuner is designed to be able to connect with third-party cloud services, such as AWS and Google Cloud, as well as users' private clouds.

\begin{figure}[!ht]
  \centering
  \includegraphics[width=1.0\linewidth]{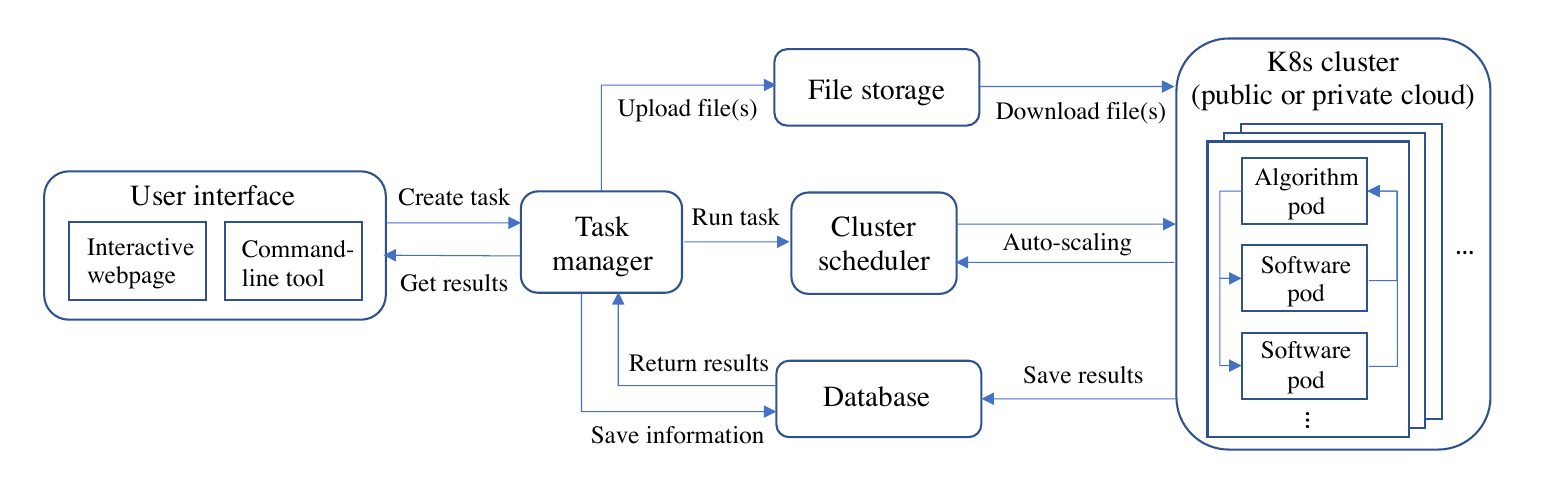}
  \caption{MindOpt Tuner's architecture with elastic cloud resource.}
  \label{fig:infra}
\end{figure}

\paragraph{Architecture} See Fig. \ref{fig:infra}. Users can create the tuning task via either the interactive webpage or the command-line tool. The task manager will upload the necessary data files to a shared file-storage and send the task execution instruction to the cluster scheduler. 
The cluster scheduler then starts a Kubernetes (K8S) pod within which the image of the algorithm and the task configurations are loaded.
Once the algorithm starts running, sub-tasks of hyperparameter evaluations will be continuously created and sent to the cluster scheduler, which will start a set of new K8S pods within which the image of software is loaded and sub-tasks are executed.
The cluster scheduler helps to automatically adjust the K8S cluster's size, up or down, based on the real-time demands, enhancing resource utilization and reducing user costs. During times with increasing task loads, a single user can simultaneously run dozens of tuning tasks, utilizing hundreds of virtual nodes. 
As the workload of users decreases, fewer pods will be maintained.
The cluster scheduler's functionality also includes managing quotas, which ensures that users' resource usage stays within specified limits.

\paragraph{Data privacy} Although cloud file storage is associated with individual user accounts and its access is protected by passwords or unique tokens, unauthorized third-party access may still occur due to vulnerable passwords and 3rd-party system hacking. For this reason, we provide an open-source, downloadable tool, called ``MindOpt Sanitizer'', which removes all the comments and identifiable information from problem-instance files. For optimization problem formats, it anonymizes the names of objectives, variables, and constraints by replacing them with generic markers OBJ, X1, X2, ..., and CON1, CON2, ..., respectively. The users can run the tool before uploading their problem instances to the cloud. When they do this, a local file that records the mapping between the original names and the generic names is saved. The file will be used to de-anonymize the solution and log files if the users wish to download and inspect them. 

\subsection{Usage of MindOpt Tuner}
In this section, we introduce three approaches for using MindOpt Tuner, including the web-based task management panel, the command-line tool, and the Python APIs\footnote{We take tuning for optimization solvers as examples throughout this section}.

\subsubsection{Web-based Task Management Panel}
MindOpt Tuner can be used via a web-based task management interface within the MindOpt Studio. The user-friendly GUI provides a convenient way for users to create, execute, monitor, and delete their tuning tasks. 

To access the management panel, users should log into the MindOpt Studio and navigate to the ``Tuning Tasks'' tab at the top of MindOpt Tuner's webpage.\footnote{https://opt.alibabacloud.com/\#/tuner/task}
\begin{figure}[!ht]
  \centering
  \includegraphics[width=1.0\linewidth]{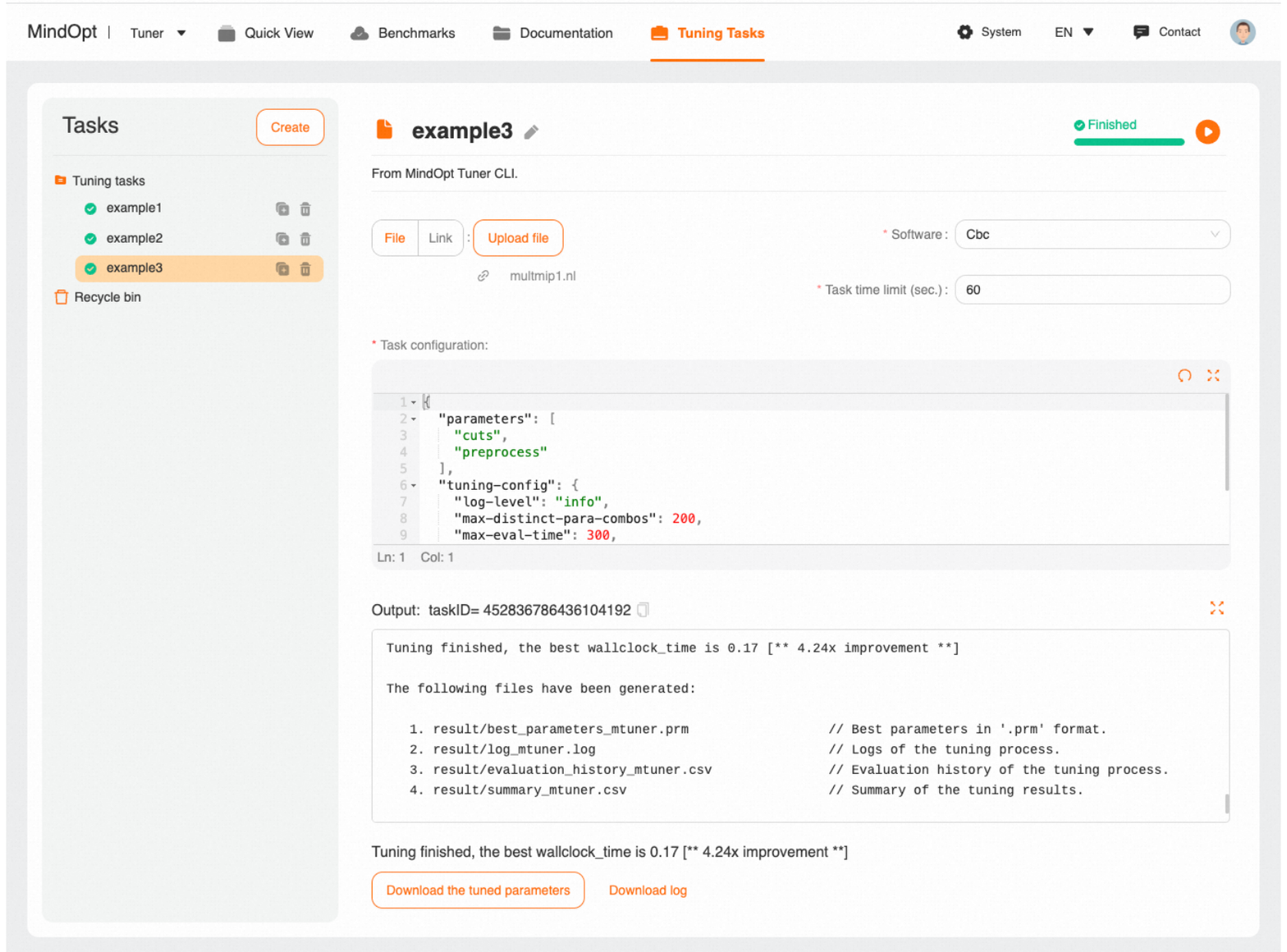}
  \caption{Screenshot of the MindOpt Tuner's task management webpage.}
  \label{fig:web_manage}
\end{figure}
Fig. \ref{fig:web_manage} displays a screenshot of the webpage: The ``Tasks'' on the left-hand side lists the created and deleted tasks respectively. 
The upper right area of the page is for uploading problem data files and specifying configurations for a tuning task. 
Simply follow the steps below to create and run a tuning task:

\noindent\textbf{Step 1: Name a new task.} To create a new task, click the ``Create'' button at the top of the ``Tasks'' and enter the task name into the pop-up window.

\noindent\textbf{Step 2: Add problem file(s).} To provide the problem data file(s) for the task, users can either select and upload local file(s) or enter the URL link(s) of the file(s) stored in Alibaba Cloud's Object Storage Service (OSS).\footnote{https://www.alibabacloud.com/product/object-storage-service}

\noindent\textbf{Step 3: Edit the task configurations.} To configure the tuning task, users need to select the target software and specify the time limit of the tuning task through the input box on the top right side. The names of the hyperparameters to tune and the additional tuning configurations should be entered in the ``Task configuration'' window using the JSON format. The basic task configurations are listed in Table \ref{tab:configs}. Please refer to the product documentation for a full list of task configurations and their valid values.

\begin{table}[!htb]
\begin{tabular}{ m{4.8cm} | m{7cm}| m{1.5cm} }
\hline
\textbf{Name of configurations} & \textbf{Description} & \textbf{Type} \\
%\midrule
\hline\hline
\verb|tuning-objective| & Specify the names of the hyperparameters to tune. & string   \\ 
\hline
\verb|max-distinct-para-combos| & Tuning task will terminate after the number of distinct hyperparameter combos reaches this number. & integer   \\ 
\hline
\verb|max-tuning-time| & Tuning task will terminate after surpassing this amount of time (in seconds).  & integer  \\
\hline
\verb|max-eval-time|  & Maximum time allowed for a single evaluation of one hyperparameter combo (in seconds). & integer  \\
 \hline 
\verb|log-level|  & Level of logging messages. & string \\  
 \hline
\verb|verbose|  & Details level of the standard output.  & integer \\
\hline
\end{tabular}
\caption{Basic task configurations of MindOpt Tuner.}\label{tab:configs}
\end{table}

\noindent\textbf{Step 4: Run the task.}
To run the configured tuning task, click the circular ``Run'' button in the upper right corner. The moving progress bar will indicate the execution progress of the task. The ``Output'' window below will display real-time outputs consisting of the intermediate information as well as the final results. 

\noindent\textbf{Step 5: Get the recommended hyperparameters.}
Upon completion of the task, the recommended values of hyperparameters will be saved to a file, which can be downloaded along with the tuning log file using the links at the bottom of the page.

\subsubsection{Command-line Tool}
The command-line tool of MindOpt Tuner provides a quick and flexible way to create and manage tuning tasks, which is especially beneficial for users who want to perform a batch of tasks in a single command or script. It can be used within MindOpt Studio Notebook\footnote{https://opt.aliyun.com/\#/platform/overview} or locally where the standalone version is installed. 

\begin{figure}[ht]
  \centering
  \includegraphics[clip,trim=0cm 0.5cm 0cm 0.5cm, width=1.1\linewidth]{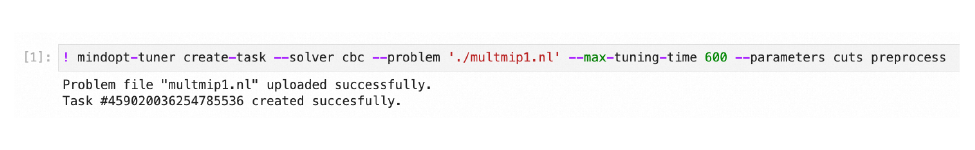}
  \caption{Example of creating a tuning task with the command-line tool.}
  \label{fig:cmd_create}
\end{figure}

For example, to create and start running a new task for tuning a solver, simply execute:
\begin{verbatim}
    mindopt-tuner create-task [-h] --solver {cbc,cplex,mindopt} --problem <problem>
\end{verbatim}
where the \verb|solver| and \verb|problem| arguments specify the target solver and the problem data file(s). Fig. \ref{fig:cmd_create} shows an example of creating a task within the MindOpt Studio Notebook.

Additional task configurations can be specified by adding arguments and values to the command. For comprehensive instructions on the usage of all commands and arguments, please refer to the product's documentation available on the website.

\subsubsection{Python APIs}
The MindOpt Tuner's Python APIs provide a set of APIs that correspond to the command-line tool and enable Python developers to seamlessly integrate the tuning process into their workflow. A quick example of using the Python APIs is provided as follows:
\begin{lstlisting}[language=Python, caption={Example of using Python APIs of MindOpt Tuner}, captionpos=b]
import mtunerpy as mtuner

args_dict = {
    'solver': 'cbc',
    'problem': 'multimlp.nl',
    'max_tuning_time': 200,
    'parameters': ['cuts', 'preprocess']
}

mtuner.create_task(args_dict)

\end{lstlisting}

\section{Numerical Experiments}\label{sec:numerical}
In this section, we present the experimental results of MindOpt Tuner's performance in tuning Cbc on the MIPLIB2017 Benchmark. The results show that MindOpt Tuner is able to achieve orders of magnitude acceleration over Cbc's default hyperparameter settings on some instances. A comparison of the performance between MindOpt Tuner and the popular tool SMAC3 reveals that our proposed tool has a clear advantage in terms of both the performance of the tuned hyperparameters and the tuning efficiency.

\subsection{Experiment Settings}
Cbc is an open-source package for solving mixed-integer linear programs. It integrates branch-and-bound, primal simplex, dual simplex, interior-point methods, etc., to find the optimal solution to a given problem. Cbc is highly customizable and can handle problems with large numbers of variables and constraints. It has been widely used in academia and industry for a variety of applications, such as transportation planning, supply chain optimization, and financial modeling.

MIPLIB2017 Benchmark set is a collection of challenging optimization problems that can be used to evaluate and compare the performance of various optimization solvers. It includes a total of 240 mixed-integer programming instances (selected from more than 1000 instances), covering a wide range of applications, such as network design, scheduling, and facility location. The benchmarks are chosen to be representative of real-world applications and have been widely used in the optimization community to benchmark solver performance and guide algorithm development.

In our tuning experiment, we focused on minimizing the \texttt{wallclock time} required for Cbc (Version: 2.10.5 Build Date: Nov 24th, 2021) to obtain the optimal solution (the solving time). We used the \textsl{speed-up ratio} as the tuning performance metric, defined as the ratio of solving time with default hyperparameters to that with tuned hyperparameters, i.e., $T_{\text{default}} / T_{\text{tuned}}$.

\subsection{Benchmarking on problems from MIPLIB2017}
\begin{figure}
    \centering
    \begin{tabular}{|c|c|c|c|c|}
    \hline
    \subf{\includegraphics[width=28mm]{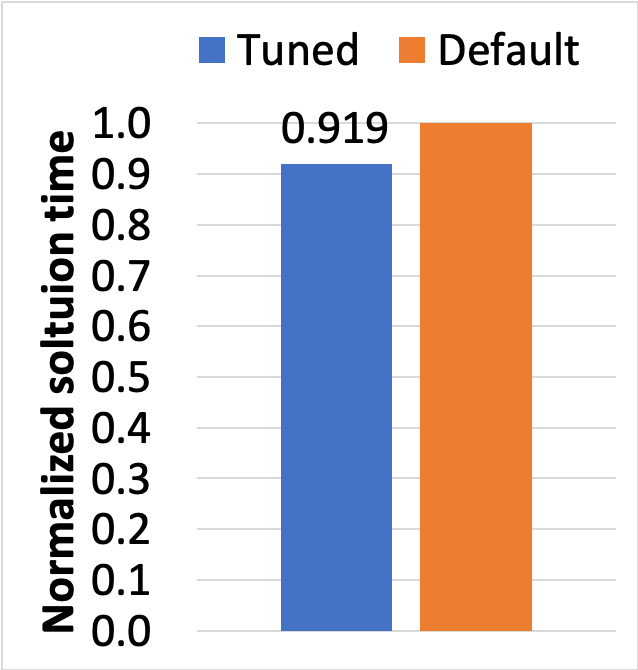}}
         {sp98ar\\(1.09x)}
    &
    \subf{\includegraphics[width=28mm]{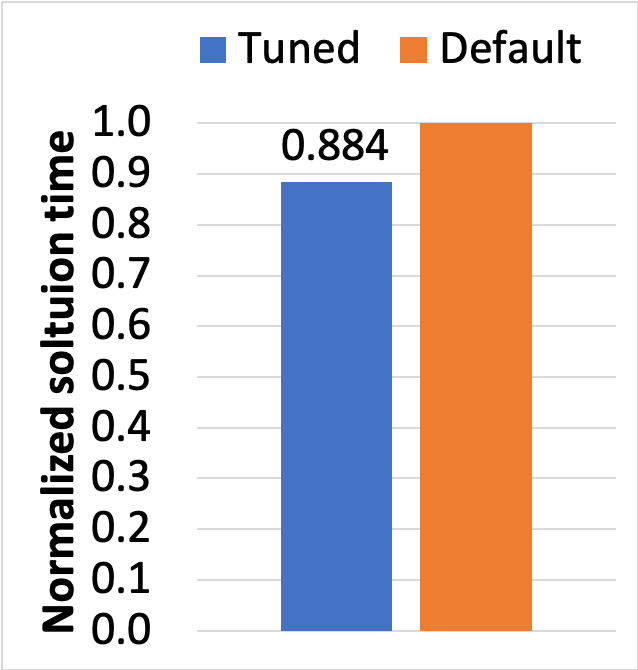}}
         {neos-4413714-turia\\(1.13x)}
    &
    \subf{\includegraphics[width=28mm]{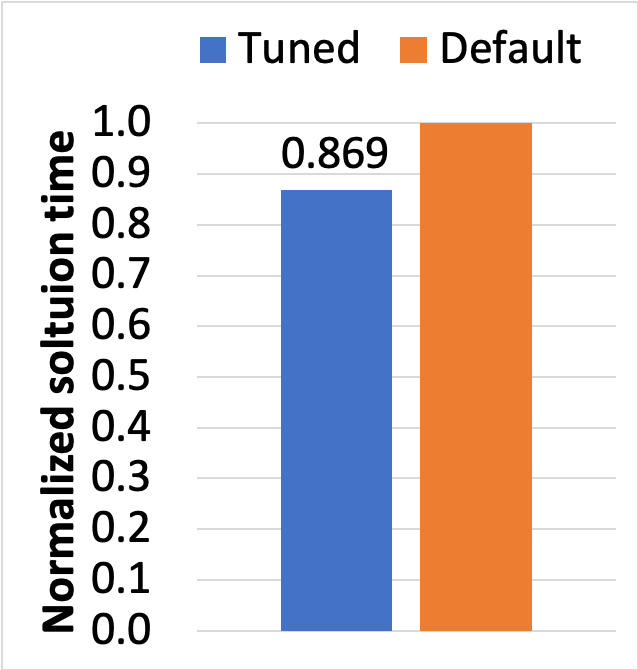}}
         {mcsched\\(1.15x)}
    &
    \subf{\includegraphics[width=28mm]{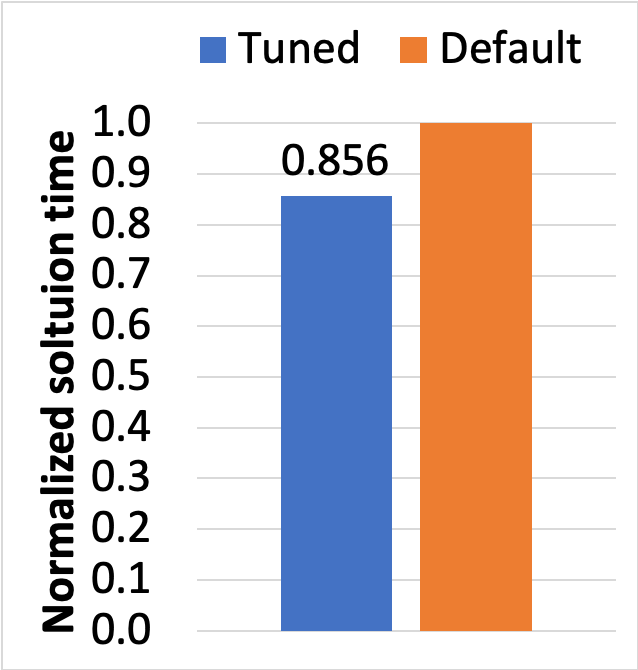}}
         {triptim1\\(1.17x)}
    &
    \subf{\includegraphics[width=28mm]{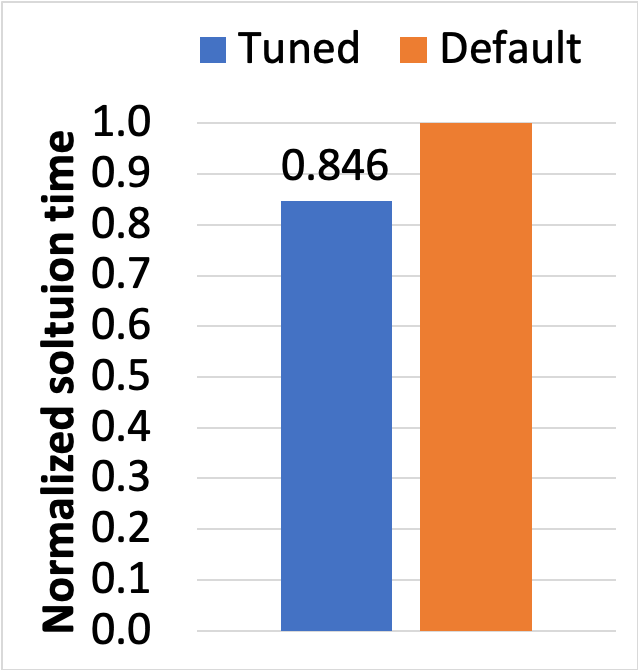}}
         {mas74\\(1.18x)}
    \\
    \hline
    \subf{\includegraphics[width=28mm]{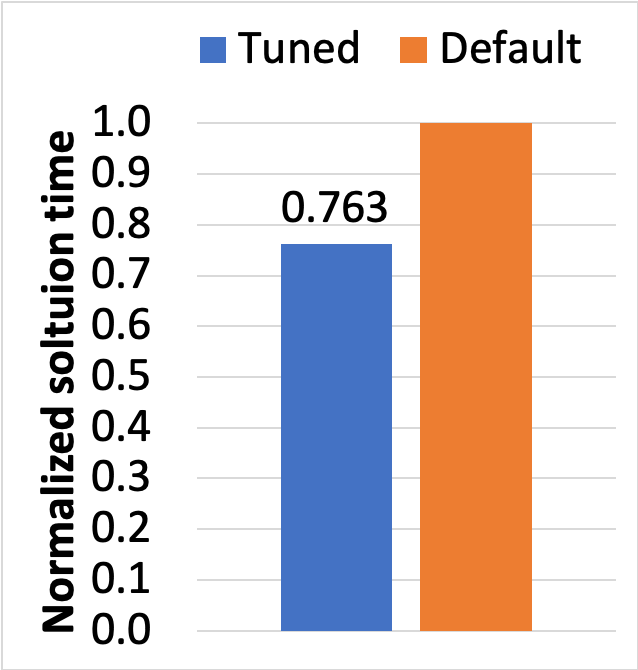}}
         {physiciansched6-2\\(1.31x)}
    &
    \subf{\includegraphics[width=28mm]{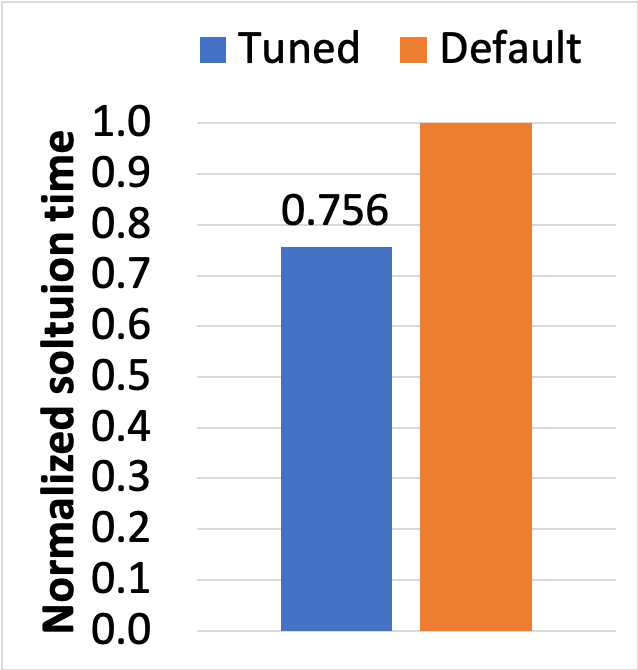}}
         {piperout-27\\(1.32x)}
    &
    \subf{\includegraphics[width=28mm]{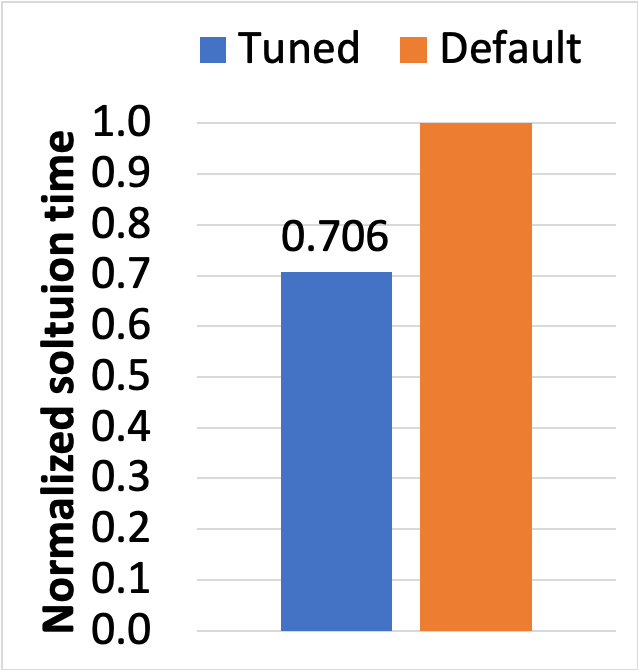}}
         {gen-ip002\\(1.42x)}
    &
    \subf{\includegraphics[width=28mm]{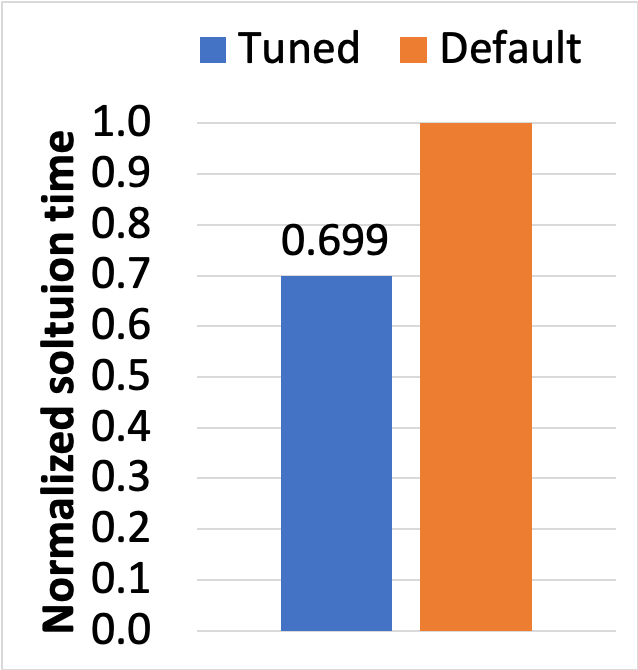}}
         {mik-250-20-75-4\\(1.43x)}
    &
    \subf{\includegraphics[width=28mm]{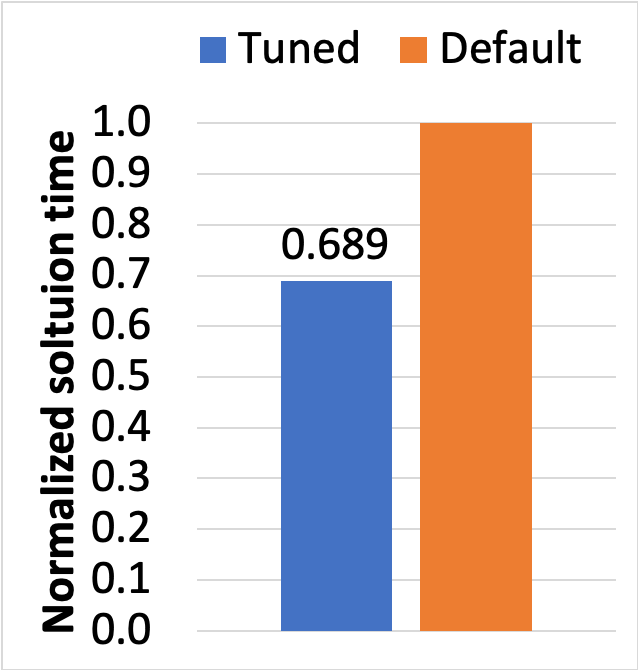}}
         {neos-2657525-crna\\(1.45x)}
    \\
    \hline
    \subf{\includegraphics[width=28mm]{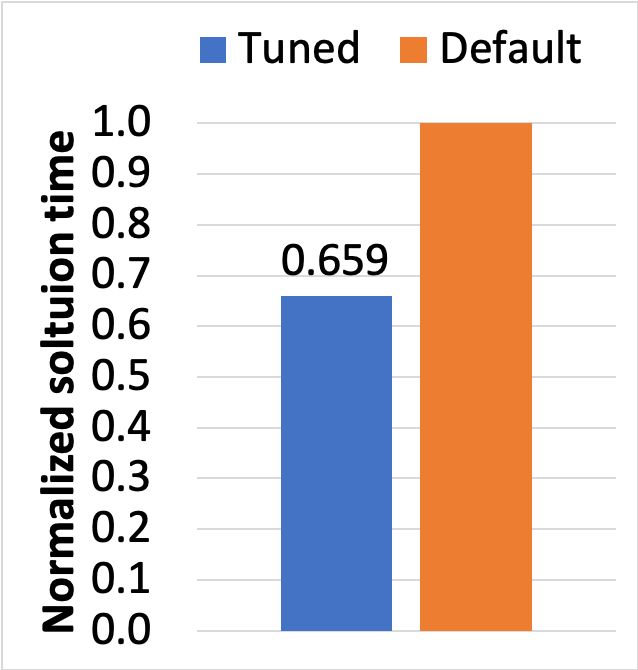}}
         {s250r10\\(1.52x)}
    &
    \subf{\includegraphics[width=28mm]{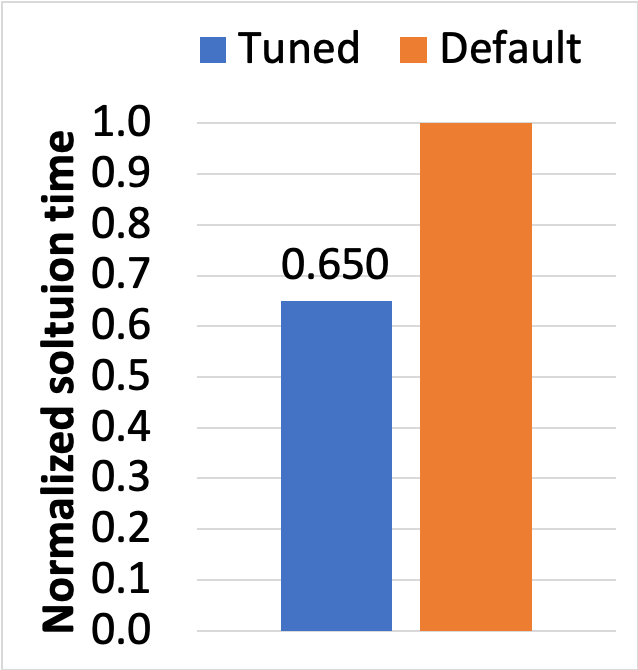}}
         {brazil3\\(1.54x)}
    &
    \subf{\includegraphics[width=28mm]{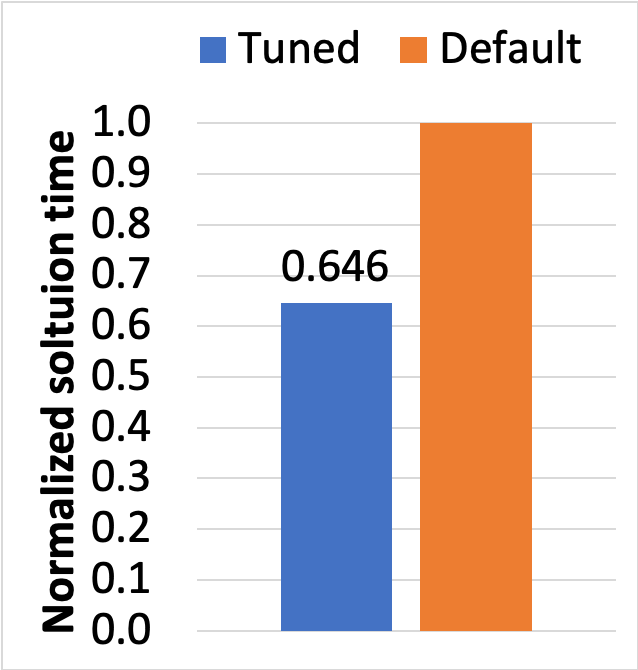}}
         {neos-1122047\\(1.55x)}
    &
    \subf{\includegraphics[width=28mm]{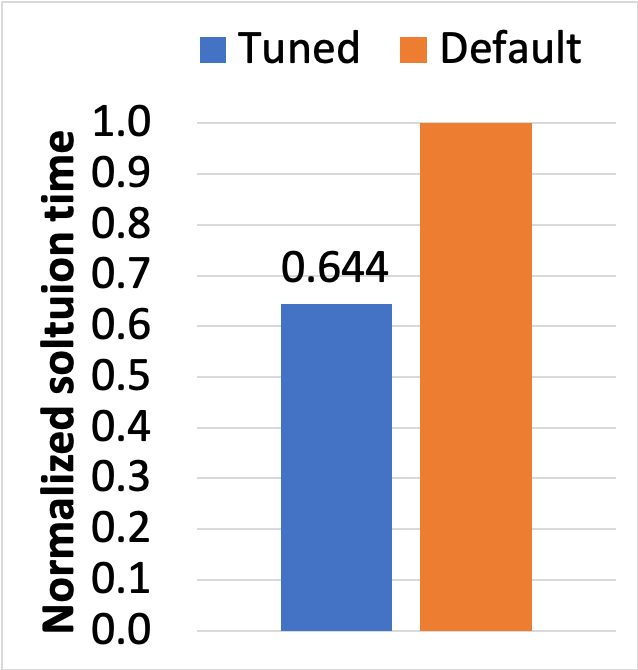}}
         {pk1\\(1.55x)}
    &
    \subf{\includegraphics[width=28mm]{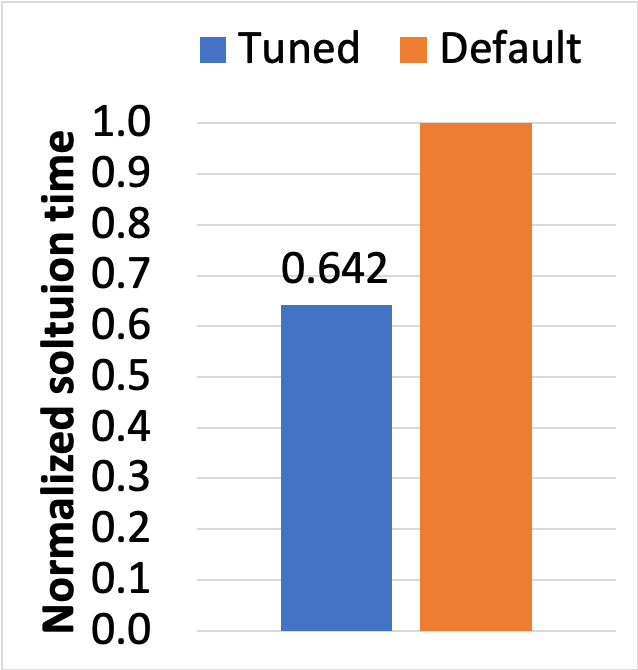}}
         {var-smallemery\\-m6j6(1.56x)}
    \\
    \hline
    \subf{\includegraphics[width=28mm]{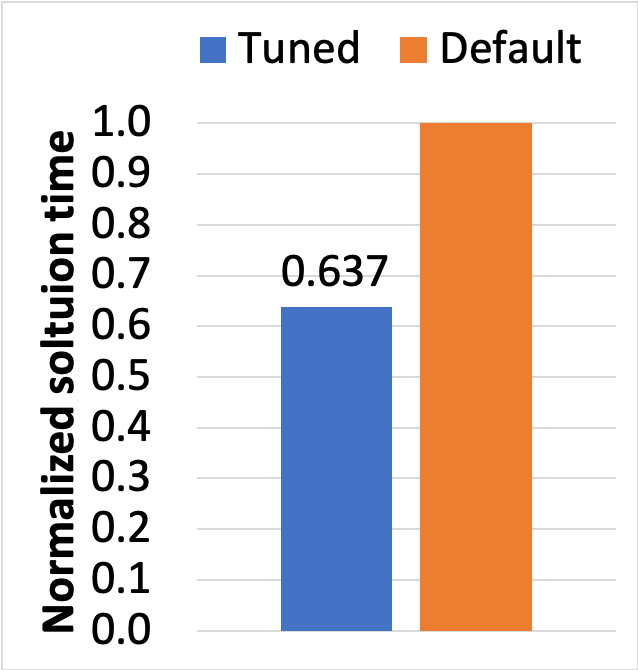}}
         {ran14x18-disj-8\\(1.57x)}
    &
    \subf{\includegraphics[width=28mm]{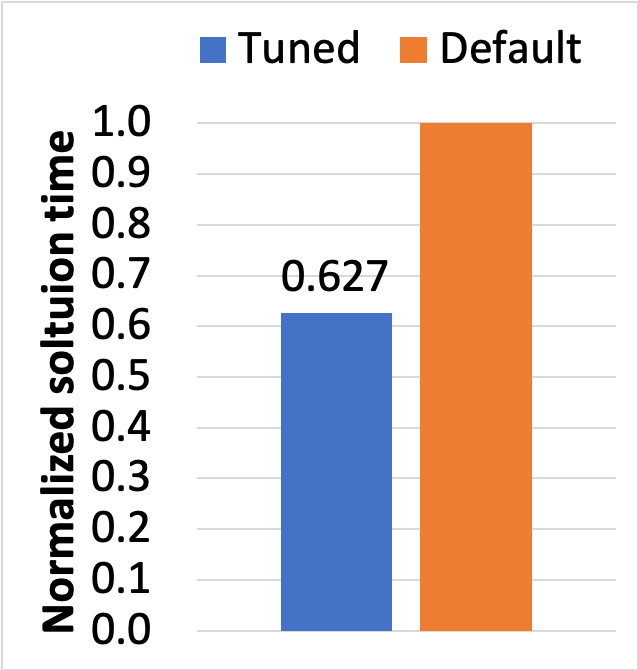}}
         {academictimetable\\small(1.6x)}
    &
    \subf{\includegraphics[width=28mm]{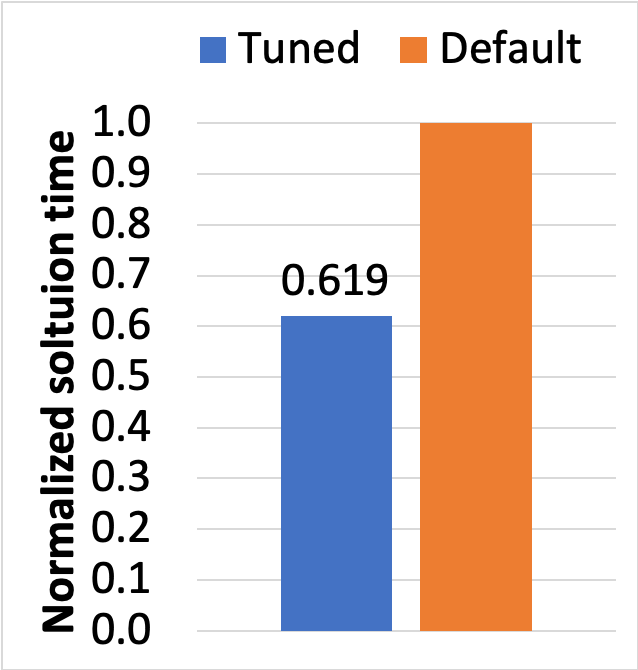}}
         {map16715-04\\(1.61x)}
    &
    \subf{\includegraphics[width=28mm]{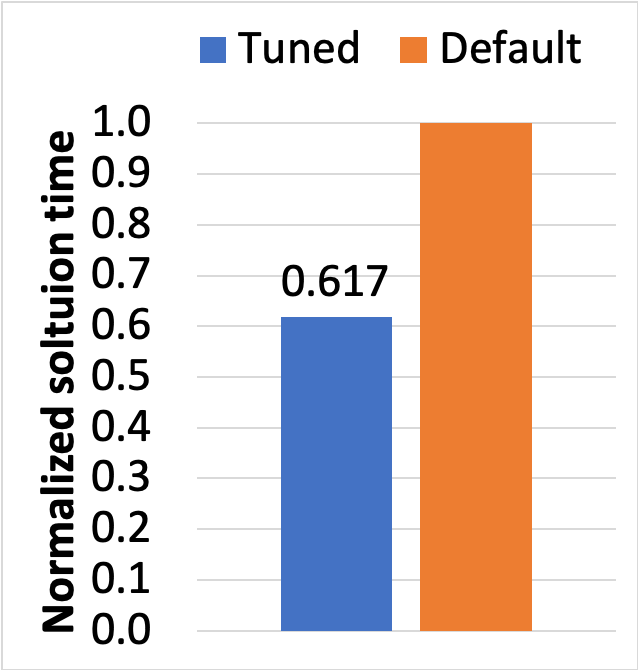}}
         {square41\\(1.62x)}
    &
    \subf{\includegraphics[width=28mm]{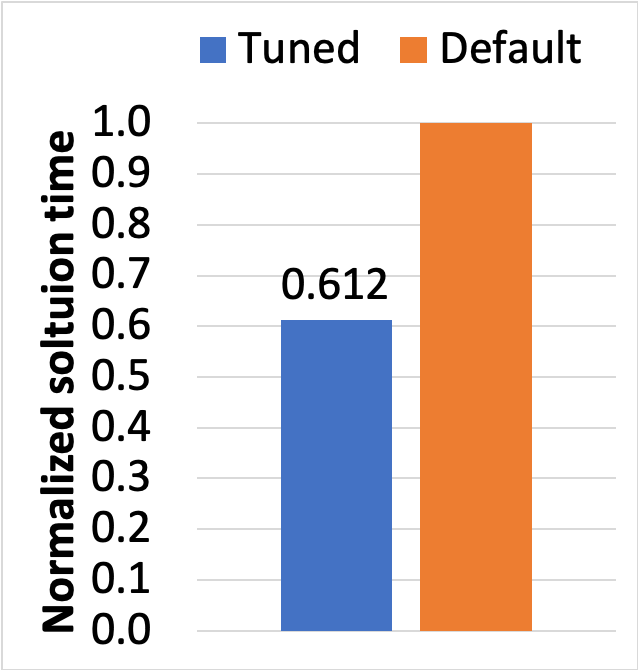}}
         {neos-3004026-krka\\(1.63x)}
    \\
    \hline
    \subf{\includegraphics[width=28mm]{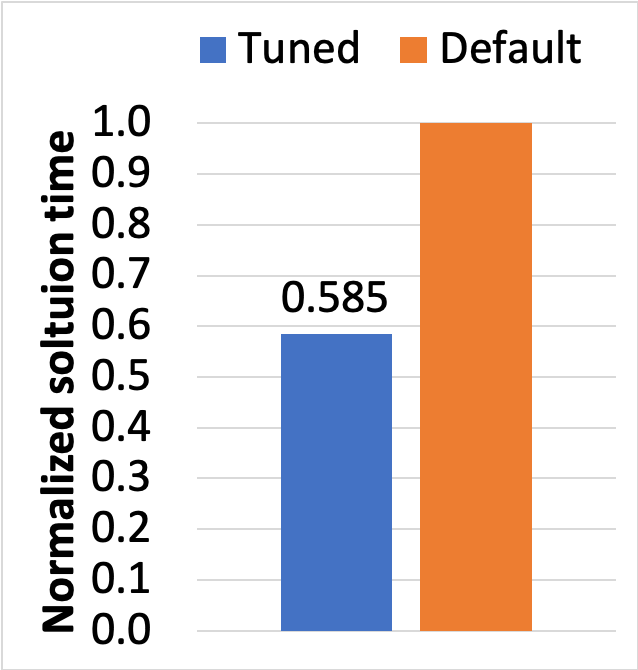}}
         {wachplan\\(1.71x)}
    &
    \subf{\includegraphics[width=28mm]{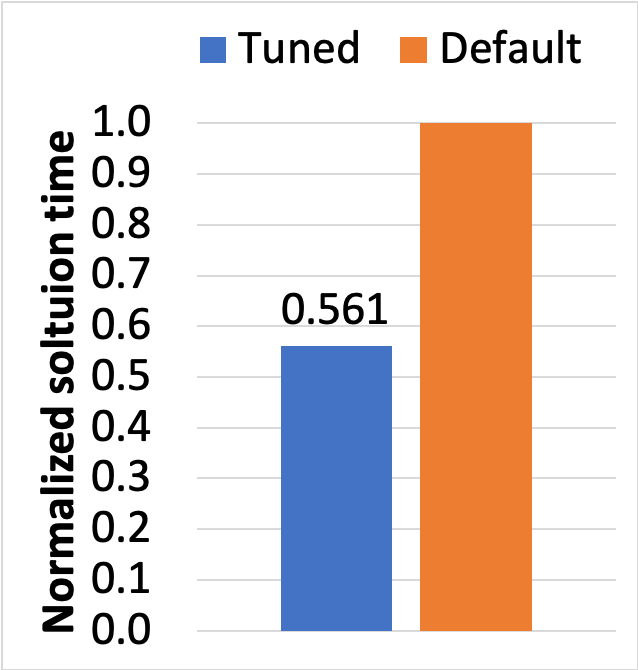}}
         {leo1\\(1.78x)}
    &
    \subf{\includegraphics[width=28mm]{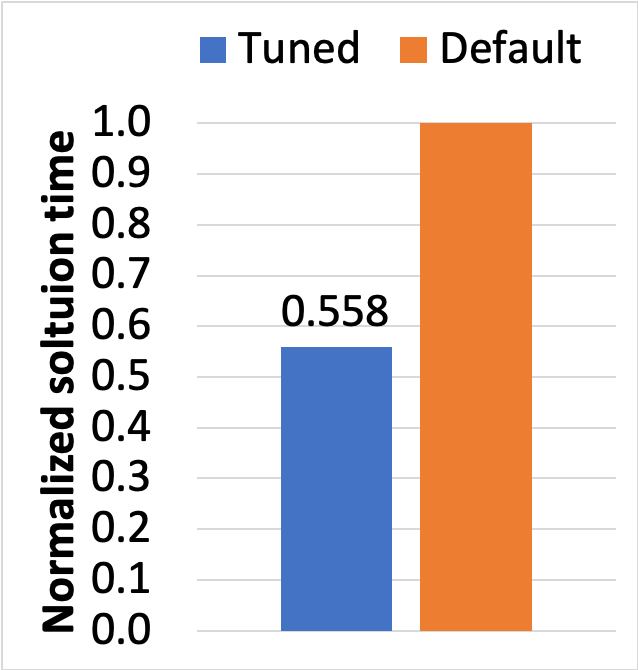}}
         {piperout-08\\(1.79x)}
    &
    \subf{\includegraphics[width=28mm]{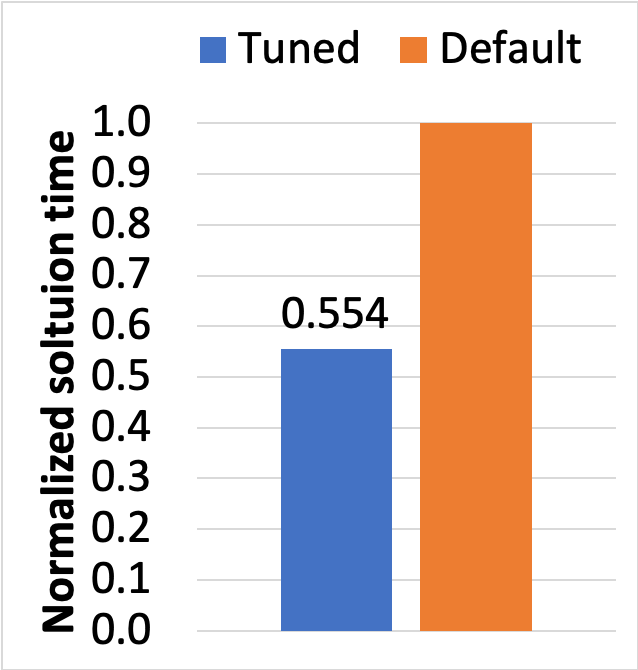}}
         {mas76\\(1.8x)}
    &
    \subf{\includegraphics[width=28mm]{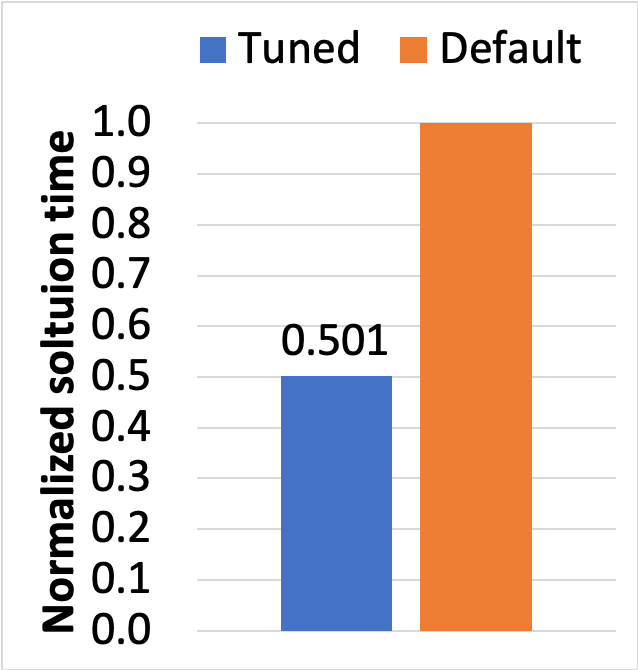}}
         {mzzv42z\\(1.99x)}
    \\
    \hline
    \end{tabular}
    \caption{MIPLIB2017 problem instances with 1x--2x speed-up ratios.}
    \centering
    \label{fig:speed2x}    
\end{figure}

\begin{figure}
    \centering
    \begin{tabular}{|c|c|c|c|}
    \hline
    \subf{\includegraphics[width=28mm]{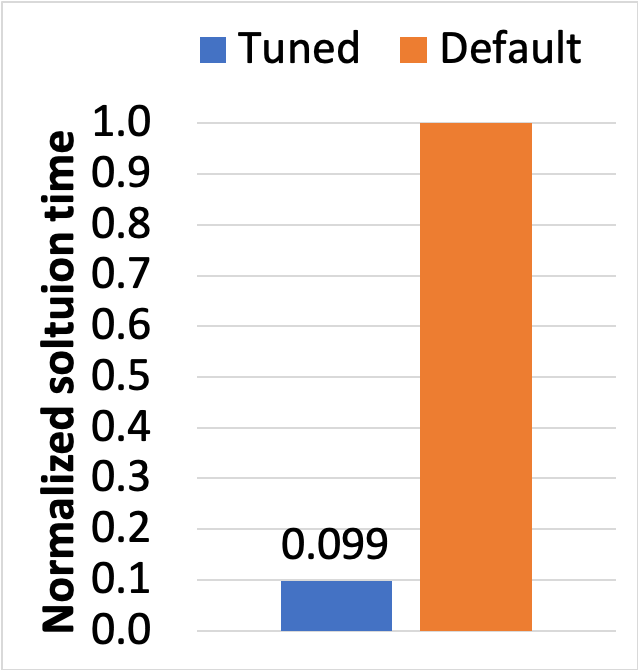}}
         {neos-3216931-puriri\\(10.13x)}
    &
    \subf{\includegraphics[width=28mm]{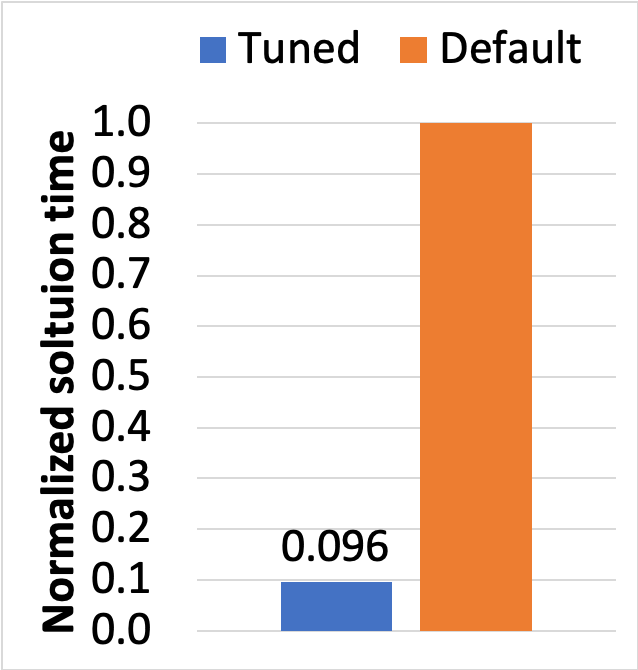}}
         {exp-1-500-5-5\\(10.45x)}
    &
    \subf{\includegraphics[width=28mm]{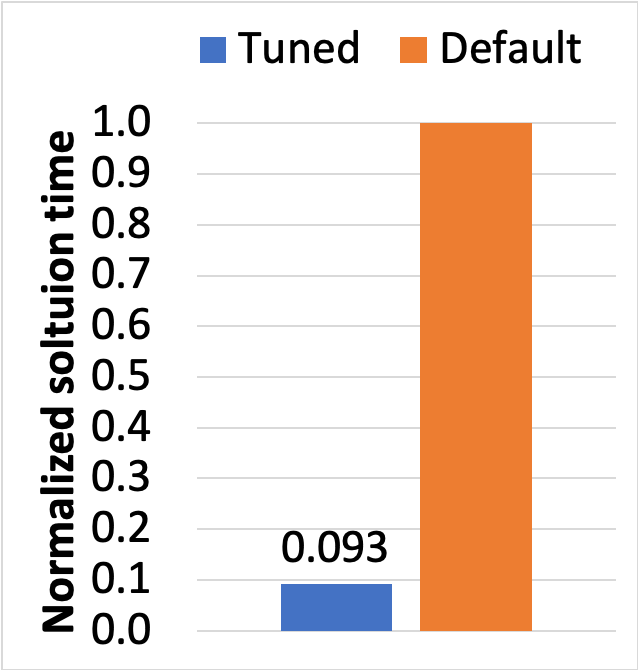}}
         {neos-1582420\\(10.78x)}
    &
    \subf{\includegraphics[width=28mm]{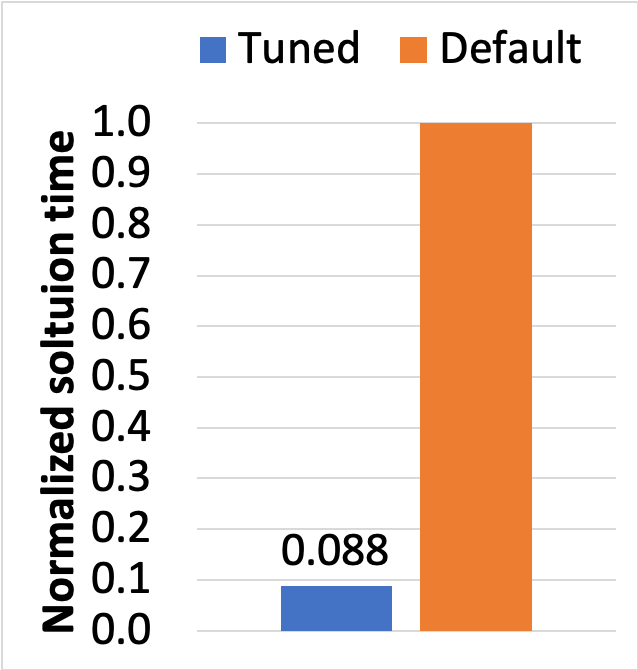}}
         {ex9\\(11.38x)}
    \\
    \hline
    \subf{\includegraphics[width=28mm]{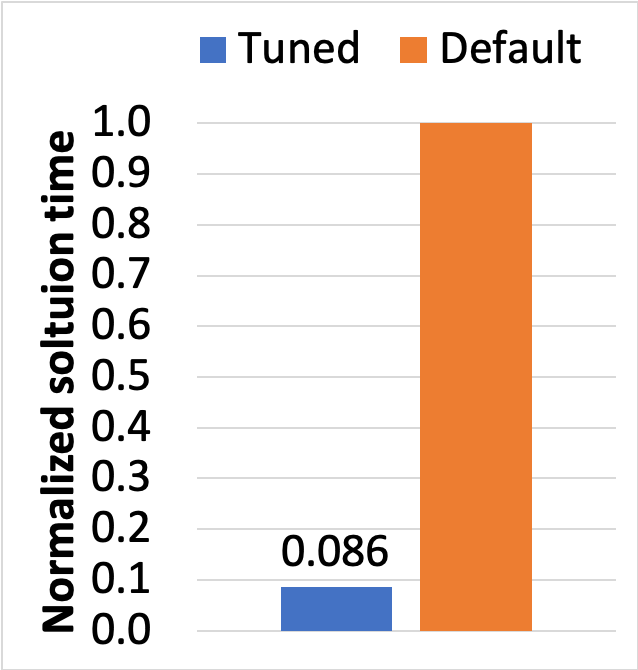}}
         {neos-5188808-nattai\\(11.69x)}
    &
    \subf{\includegraphics[width=28mm]{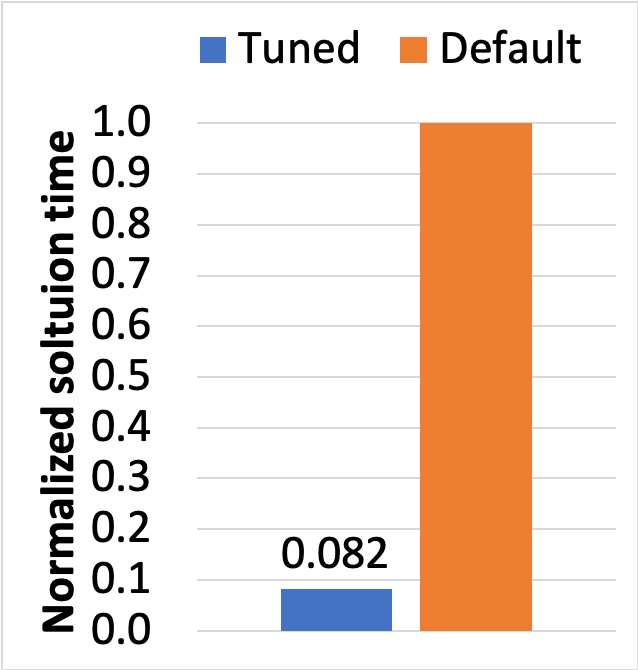}}
         {drayage-100-23\\(12.21x)}
    &
    \subf{\includegraphics[width=28mm]{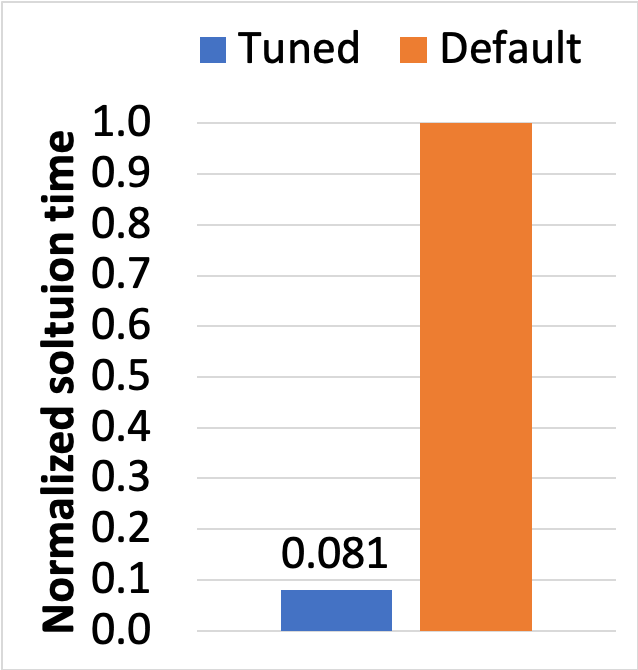}}
         {fiball\\(12.37x)}
    &
    \subf{\includegraphics[width=28mm]{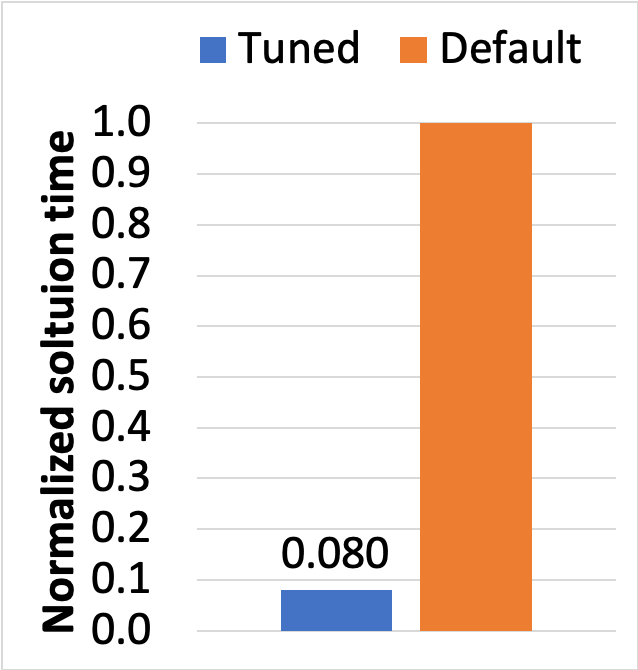}}
         {irp\\(12.45x)}
    \\
    \hline
    \subf{\includegraphics[width=28mm]{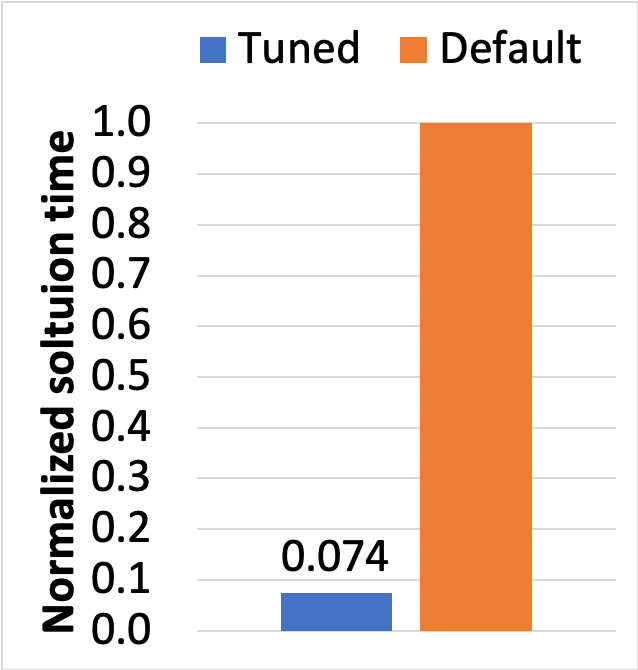}}
         {decomp2\\(13.44x)}
    &
    \subf{\includegraphics[width=28mm]{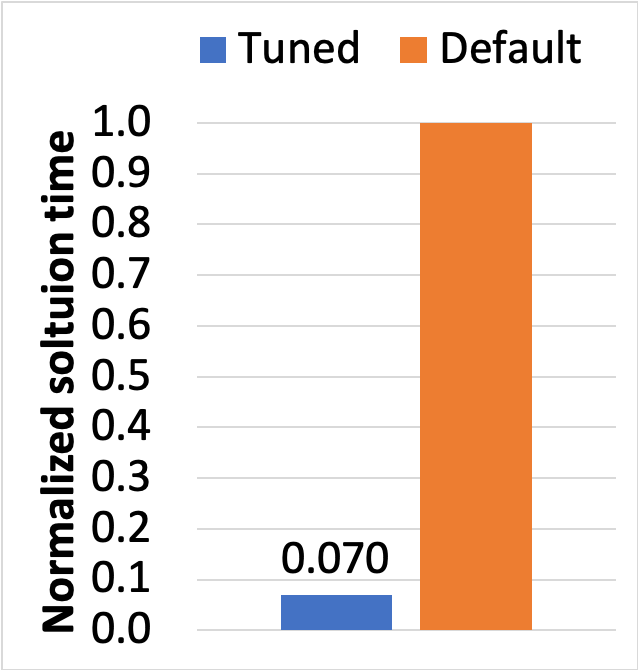}}
         {hypothyroid-k1\\(14.36x)}
    &
    \subf{\includegraphics[width=28mm]{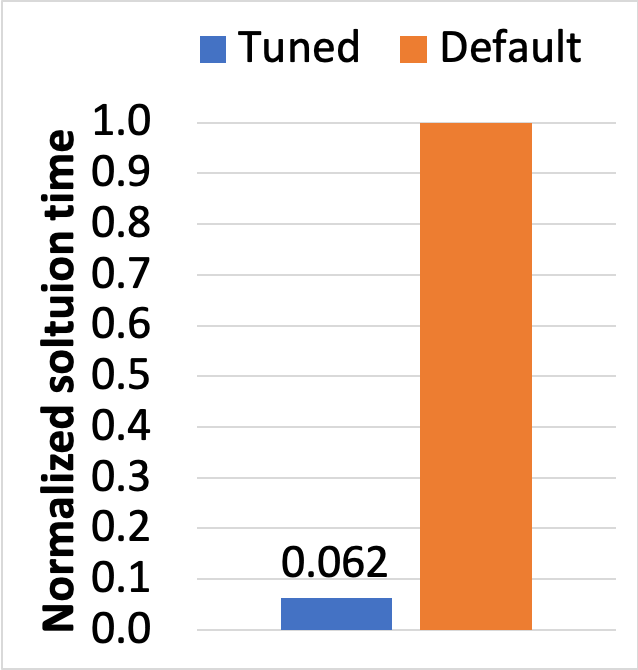}}
         {supportcase26\\(16.12x)}
    &
    \subf{\includegraphics[width=28mm]{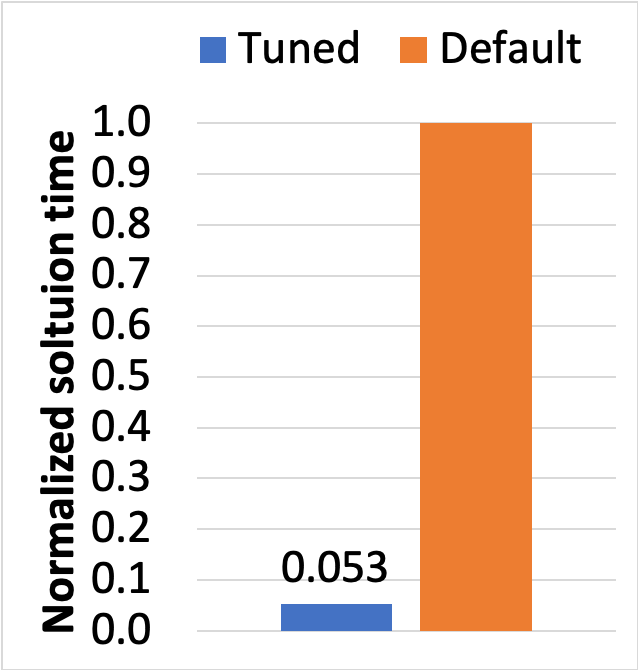}}
         {blp-ic98\\(18.94x)}
    \\
    \hline
    \subf{\includegraphics[width=28mm]{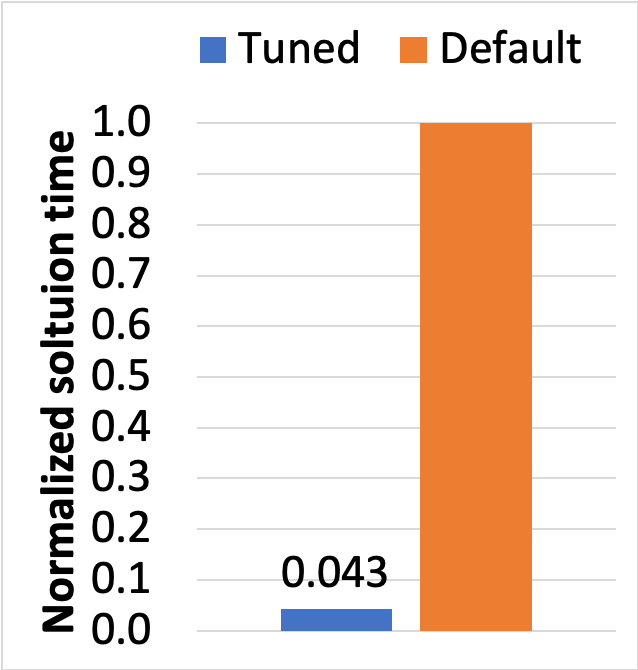}}
         {neos-1354092\\(23.35x)}
    &
    \subf{\includegraphics[width=28mm]{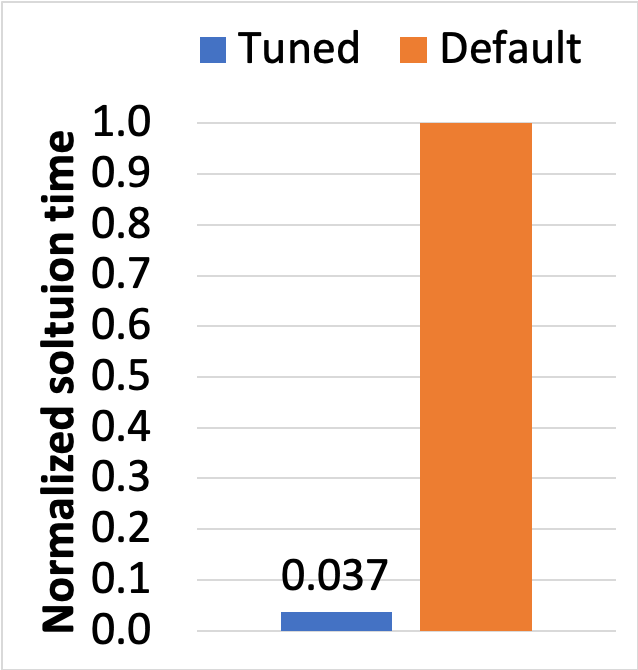}}
         {neos-950242\\(27.07x)}
    &
    \subf{\includegraphics[width=28mm]{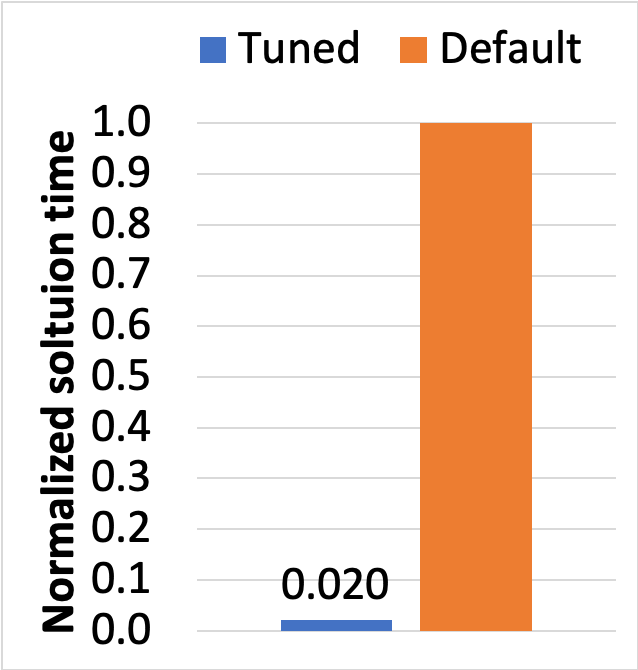}}
         {neos17\\-m6j6(49.23x)}
    &
    \subf{\includegraphics[width=28mm]{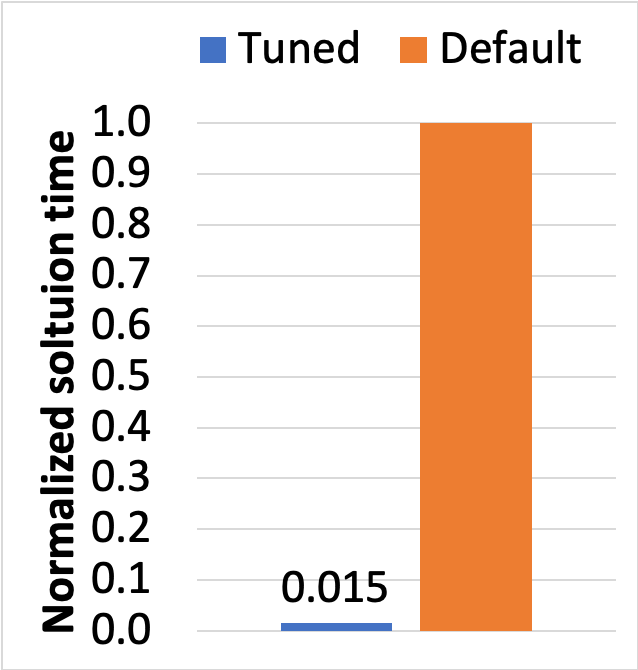}}
         {n2seq36q\\small(68.7x)}
    \\
    \hline
    \end{tabular}
    \caption{MIPLIB2017 problem instances with 10x--100x speed-up ratio.}
    \centering
    \label{fig:speed100x}    
\end{figure}

\begin{figure}[htb]
  \centering
  \includegraphics[width=0.93\linewidth]{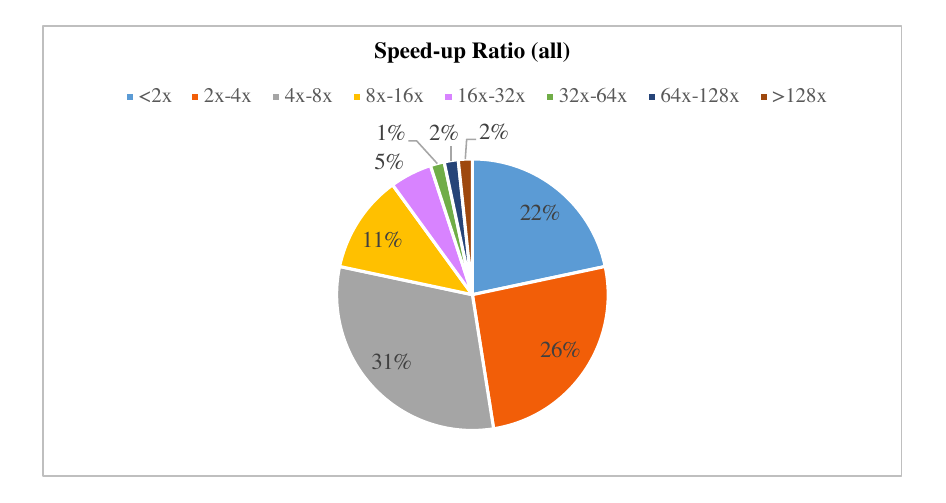}
  \caption{Distribution of speed-up ratios.}
  \label{fig:speed_up_pie}
\end{figure}

In the first set of experiments, the tuning task for each individual problem instance was conducted on a Linux machine with a 2.50 GHz Intel(R) Xeon(R) Platinum 8163 CPU, 16 cores (2 threads per core), and 30GB memory. The results demonstrate that the tuned parameters recommended by MindOpt Tuner are effective in enhancing the performance of Cbc. Fig. \ref{fig:speed2x} and Fig. \ref{fig:speed100x} show the speed-up ratio of two groups of problem instances: those with solving times ranging between 2000s and 12000s (high-difficulty problems) and those with solving times ranging between 500s and 2000s (moderate-difficulty problems). The distribution of the achieved speed-up ratios is displayed in Fig. \ref{fig:speed_up_pie}. It is clear that, for over three-quarters of the problems considered, a speed-up ratio greater than 2x (equivalent to at least a 50\% reduction in solving time) can be achieved. Additionally, more than half of the problems considered can achieve a speed-up ratio greater than 4x. Moreover, 10\% of the considered problems have achieved a speed-up ratio exceeding 32x, including 2\% of the problems that have achieved a speed-up ratio of 100x or more (with the maximum speed-up ratio exceeding 1000).

\begin{figure}[!t]
  \centering
  \includegraphics[width=0.98\linewidth]{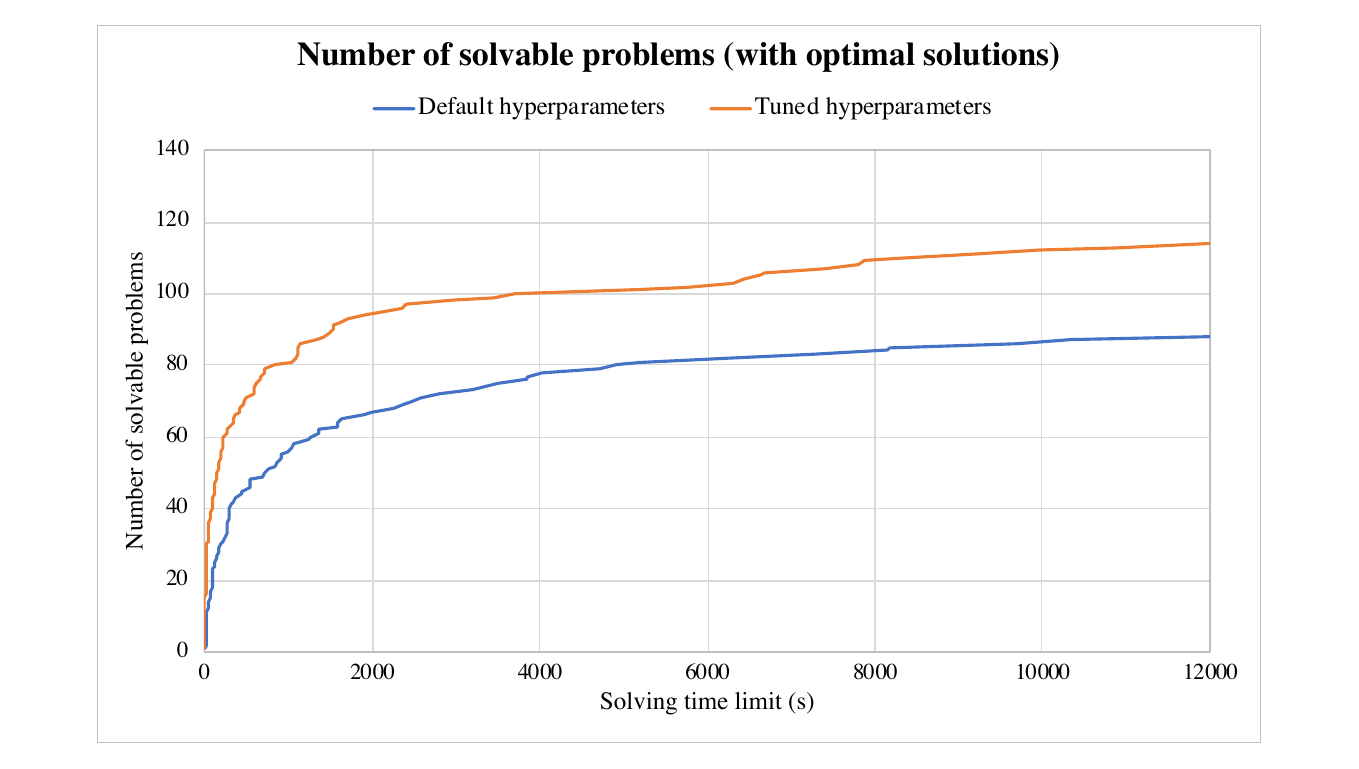}
  \caption{Number of solvable problems vs solving time limit in seconds.}
  \label{fig:solved_prob}
\end{figure}

We have also analyzed the number of problems that can be solved within different time budgets before and after using the parameters recommended by MindOpt Tuner. As illustrated in Fig. \ref{fig:solved_prob}, the number of solvable problems with the tuned parameters is substantially greater than that with default parameters at a given solving time limit. For instance, only 77 problems can be solved with default parameters in 4000s. While with the tuned parameters, an additional 23 problems can be solved within the same time limit.

\subsubsection{Performance Comparison with SMAC3}
Based on the tuning results of the MIPLIB2017 instances, we selected a set of problem instances that varied in their solving time under default hyperparameters and their achieved speed-up ratios with MindOpt Tuner. We then tuned these problems using both MindOpt Tuner and SMAC3 (V1.4) with their default settings and under the same limit on hyperparameter evaluations (200 per instance). The tuning performance of the two tools was then compared in terms of the speed-up ratio and elapsed tuning time\footnote{We did not choose ParamILS as the baseline because it is an earlier work from the same group of SMAC3, with some of its designs already included in SMAC3. \emph{irace} is not chosen here because it is designed particularly for the scenario where running time is not the tuning objective \cite{irace2016}, which is conflict with our experimental settings.}.

\begin{table}[!]
\centering
% \begin{tabular}{ c{2cm} | c{2cm}| c{3cm} | c{2cm} | c{2cm} }
% \hline
\begin{tabular}{c|c|c|cc}
\toprule
\textbf{Instance}  & \textbf{Default solution time}      & \textbf{Tool} & \multicolumn{1}{c|}{\textbf{Speed-up ratio}} & \textbf{Tuning time}               \\ \cline{1-5} 
\textbf{app1-1.mps} & 29.23s     & SMAC3                           & \multicolumn{1}{c|}{6.83x}           & 0.45h          \\ \cline{3-5} 
                          &                & MindOpt Tuner                   & \multicolumn{1}{c|}{\textbf{8.61x}}           & \textbf{0.21h}        \\ \hline
\textbf{air05.mps} & 125.57s      & SMAC3                           & \multicolumn{1}{c|}{3.11x}           & 5.85h          \\ \cline{3-5} 
                          &                & MindOpt Tuner                   & \multicolumn{1}{c|}{\textbf{3.23x}}           & \textbf{1.67h}          \\ \hline
\textbf{neos8.mps} & 108.82s      & SMAC3                           & \multicolumn{1}{c|}{3.38x}           & 7.32h          \\ \cline{3-5} 
                          &                & MindOpt Tuner                   & \multicolumn{1}{c|}{3.16x}           & 1.31h          \\ \hline
\textbf{nu25-pr12.mps}    & 185.63s     & SMAC3                           & \multicolumn{1}{c|}{7.18x}           & 5.1h        \\ \cline{3-5} 
                          &                & MindOpt Tuner                   & \multicolumn{1}{c|}{\textbf{8.07x}}           & \textbf{1.99h}          \\ \hline                          
\textbf{eil33-2.mps} & 472.24s      & SMAC3                           & \multicolumn{1}{c|}{24.73x}           & 15.13h          \\ \cline{3-5} 
                          &                & MindOpt Tuner                   & \multicolumn{1}{c|}{24.12x}           & 2.37h          \\ \hline                           
\textbf{swath3.mps}  & 427.87s    & SMAC3                           & \multicolumn{1}{c|}{2.17x}           & 20.05h         \\ \cline{3-5} 
                         &                 & MindOpt Tuner                   & \multicolumn{1}{c|}{\textbf{3.43x}}           & \textbf{5.56h}          \\ \bottomrule
\end{tabular}
\caption{Tuning performance comparison between SMAC3 and MindOpt Tuner.}
\label{tab:my-table}
\end{table}

The selected problems have default solution times ranging from less than 30s to more than 400s, and the speed-up ratios by MindOpt Tuner vary from less than 2x to more than 10x. As shown in Table \ref{tab:my-table}, MindOpt Tuner outperforms SMAC3 in terms of the speed-up ratios. Additionally, MindOpt Tuner generally requires less tuning time compared to SMAC3, and this advantage becomes more significant as the difficulty of the problem increases.

\section{Conclusions}\label{sec:conclusion}
In this work, we present MindOpt Tuner, a cloud-based efficient hyperparameter tuning tool designed for numerical software. We explain the architecture design that enables the utilization of elastic cloud resources and several of its benefits. We also introduce the three approaches: the webpage-based interface, the command-line tool, and the Python APIs by which users can access MindOpt Tuner. Finally, we provide numerical experiment results of tuning Cbc's hyperparameters on the MIPLIB benchmark, which validates the tuning performance of MindOpt Tuner as well as its efficiency advantages compared to SMAC3.

We are continually developing new features that enrich the current functionality of MindOpt Tuner. For example, future versions will enable saving the state of a tuning task as a checkpoint, which can be loaded whenever users want to continue this tuning task.
In addition, to enhance the generalization ability of MindOpt Tuner, we are developing a hyperparameter recommendation module that can directly recommend hyperparameter for newly unseen problem data.

\begin{acknowledgements}
This work of M. Zhang and M. Wang is supported in part by the National Key Research and Development Program of China (Grant No. 2022YFB2403500).
\end{acknowledgements}

% \bibliography{example}
\bibliography{mindopt_tuner}

\begin{thebibliography}{26}
\providecommand{\natexlab}[1]{#1}
\providecommand{\url}[1]{\texttt{#1}}
\expandafter\ifx\csname urlstyle\endcsname\relax
  \providecommand{\doi}[1]{doi: #1}\else
  \providecommand{\doi}{doi: \begingroup \urlstyle{rm}\Url}\fi

\bibitem[Akiba et~al.(2019)Akiba, Sano, Yanase, Ohta, and Koyama]{optuna2019}
Akiba, T., Sano, S., Yanase, T., Ohta, T., and Koyama, M.
\newblock Optuna: A next-generation hyperparameter optimization framework.
\newblock In \emph{Proceedings of the 25th {ACM} {SIGKDD} International
  Conference on Knowledge Discovery and Data Mining}, 2019.

\bibitem[Bergstra \& Bengio(2012)Bergstra and Bengio]{random_search}
Bergstra, J. and Bengio, Y.
\newblock Random search for hyper-parameter optimization.
\newblock \emph{J. Mach. Learn. Res.}, 13\penalty0 (null):\penalty0 281–305,
  feb 2012.

\bibitem[Bergstra et~al.(2013)Bergstra, Yamins, and Cox]{hyperopt2013}
Bergstra, J., Yamins, D., and Cox, D.~D.
\newblock Making a science of model search: Hyperparameter optimization in
  hundreds of dimensions for vision architectures.
\newblock In \emph{International Conference on Machine Learning}, 2013.

\bibitem[Clark \& Hayes(2019)Clark and Hayes]{sigopt}
Clark, S. and Hayes, P.
\newblock {SigOpt} {W}eb page, 2019.
\newblock URL \url{https://sigopt.com}.

\bibitem[Cowen-Rivers et~al.(2022)Cowen-Rivers, Lyu, Tutunov, Wang, Grosnit,
  Griffiths, Maravel, Hao, Wang, Peters, and Bou~Ammar]{hebo2022}
Cowen-Rivers, A., Lyu, W., Tutunov, R., Wang, Z., Grosnit, A., Griffiths,
  R.-R., Maravel, A., Hao, J., Wang, J., Peters, J., and Bou~Ammar, H.
\newblock Hebo: Pushing the limits of sample-efficient hyperparameter
  optimisation.
\newblock \emph{Journal of Artificial Intelligence Research}, 74, 07 2022.

\bibitem[Cplex(2009)]{cplex2009v12}
Cplex, I.~I.
\newblock V12. 1: User’s manual for cplex.
\newblock \emph{International Business Machines Corporation}, 46\penalty0
  (53):\penalty0 157, 2009.

\bibitem[Falkner et~al.(2018)Falkner, Klein, and Hutter]{bohb2018}
Falkner, S., Klein, A., and Hutter, F.
\newblock Bohb: Robust and efficient hyperparameter optimization at scale.
\newblock In \emph{International Conference on Machine Learning}, 2018.

\bibitem[Forrest \& Lougee-Heimer(2005)Forrest and Lougee-Heimer]{cbc}
Forrest, J. and Lougee-Heimer, R.
\newblock \emph{CBC User Guide}, chapter Chapter 10, pp.\  257--277.
\newblock INFORMS, 2005.

\bibitem[Ge et~al.(2022)Ge, Huangfu, Wang, Wu, and Ye]{copt}
Ge, D., Huangfu, Q., Wang, Z., Wu, J., and Ye, Y.
\newblock Cardinal {O}ptimizer {(COPT)} user guide.
\newblock https://guide.coap.online/copt/en-doc, 2022.

\bibitem[Gleixner et~al.(2021)Gleixner, Hendel, Gamrath, Achterberg, Bastubbe,
  Berthold, Christophel, Jarck, Koch, Linderoth, L\"ubbecke, Mittelmann,
  Ozyurt, Ralphs, Salvagnin, and Shinano]{miplib2017}
Gleixner, A., Hendel, G., Gamrath, G., Achterberg, T., Bastubbe, M., Berthold,
  T., Christophel, P.~M., Jarck, K., Koch, T., Linderoth, J., L\"ubbecke, M.,
  Mittelmann, H.~D., Ozyurt, D., Ralphs, T.~K., Salvagnin, D., and Shinano, Y.
\newblock {MIPLIB 2017: Data-Driven Compilation of the 6th Mixed-Integer
  Programming Library}.
\newblock \emph{Mathematical Programming Computation}, 2021.

\bibitem[Golovin et~al.(2017)Golovin, Solnik, Moitra, Kochanski, Karro, and
  Sculley]{vizier2017}
Golovin, D., Solnik, B., Moitra, S., Kochanski, G., Karro, J., and Sculley, D.
\newblock Google vizier: A service for black-box optimization.
\newblock In \emph{Proceedings of the 23rd ACM SIGKDD International Conference
  on Knowledge Discovery and Data Mining}, pp.\  1487–1495, 2017.

\bibitem[{Gurobi Optimization, Inc.}(2023)]{gurobi}
{Gurobi Optimization, Inc.}
\newblock Gurobi optimizer reference manual, 2023.
\newblock URL \url{http://www.gurobi.com}.

\bibitem[Hansen et~al.(2003)Hansen, M{\"{u}}ller, and Koumoutsakos]{cma2003}
Hansen, N., M{\"{u}}ller, S.~D., and Koumoutsakos, P.
\newblock Reducing the time complexity of the derandomized evolution strategy
  with covariance matrix adaptation {(CMA-ES)}.
\newblock \emph{Evol. Comput.}, 11\penalty0 (1):\penalty0 1--18, 2003.

\bibitem[Hutter et~al.(2009)Hutter, Hoos, Leyton-Brown, and
  St\"{u}tzle"]{ParamILS}
Hutter, F., Hoos, H.~H., Leyton-Brown, K., and St\"{u}tzle", T.
\newblock Paramils: An automatic algorithm configuration framework.
\newblock \emph{Journal of Artificial Intelligence Research}, 36:\penalty0
  267--306, October 2009.

\bibitem[Hutter et~al.(2019)Hutter, Kotthoff, and
  Vanschoren]{hutter2019automated}
Hutter, F., Kotthoff, L., and Vanschoren, J. (eds.).
\newblock \emph{Automated Machine Learning - Methods, Systems, Challenges}.
\newblock Springer, 2019.

\bibitem[Kandasamy et~al.(2020)Kandasamy, Vysyaraju, Neiswanger, Paria,
  Collins, Schneider, Poczos, and Xing]{dragonfly2020}
Kandasamy, K., Vysyaraju, K.~R., Neiswanger, W., Paria, B., Collins, C.~R.,
  Schneider, J., Poczos, B., and Xing, E.~P.
\newblock Tuning hyperparameters without grad students: Scalable and robust
  bayesian optimisation with dragonfly.
\newblock \emph{Journal of Machine Learning Research}, 21\penalty0
  (81):\penalty0 1--27, 2020.

\bibitem[Liaw et~al.(2018)Liaw, Liang, Nishihara, Moritz, Gonzalez, and
  Stoica]{raytune2018}
Liaw, R., Liang, E., Nishihara, R., Moritz, P., Gonzalez, J.~E., and Stoica, I.
\newblock Tune: A research platform for distributed model selection and
  training.
\newblock \emph{arXiv preprint arXiv:1807.05118}, 2018.

\bibitem[Lindauer et~al.(2022)Lindauer, Eggensperger, Feurer, Biedenkapp, Deng,
  Benjamins, Ruhkopf, Sass, and Hutter]{smac2022}
Lindauer, M., Eggensperger, K., Feurer, M., Biedenkapp, A., Deng, D.,
  Benjamins, C., Ruhkopf, T., Sass, R., and Hutter, F.
\newblock Smac3: A versatile bayesian optimization package for hyperparameter
  optimization.
\newblock \emph{Journal of Machine Learning Research}, 23\penalty0
  (54):\penalty0 1--9, 2022.

\bibitem[Liu et~al.(2017)Liu, Hu, Qian, Yu, and Qian]{zoopt2017}
Liu, Y.-R., Hu, Y.-Q., Qian, H., Yu, Y., and Qian, C.
\newblock Zoopt: a toolbox for derivative-free optimization.
\newblock \emph{Science China Information Sciences}, 65, 2017.

\bibitem[López-Ibáñez et~al.(2016)López-Ibáñez, Dubois-Lacoste, {Pérez
  Cáceres}, Birattari, and Stützle]{irace2016}
López-Ibáñez, M., Dubois-Lacoste, J., {Pérez Cáceres}, L., Birattari, M.,
  and Stützle, T.
\newblock The irace package: Iterated racing for automatic algorithm
  configuration.
\newblock \emph{Operations Research Perspectives}, 3:\penalty0 43--58, 2016.

\bibitem[Montgomery(2017)]{montgomery2017design}
Montgomery, D.
\newblock \emph{Design and Analysis of Experiments}.
\newblock Wiley, 2017.

\bibitem[Rakotoarison et~al.(2019)Rakotoarison, Schoenauer, and
  Sebag]{mosaic2019}
Rakotoarison, H., Schoenauer, M., and Sebag, M.
\newblock Automated machine learning with monte-carlo tree search.
\newblock In \emph{Proceedings of the Twenty-Eighth International Joint
  Conference on Artificial Intelligence, {IJCAI-19}}, pp.\  3296--3303, 7 2019.

\bibitem[Rapin \& Teytaud(2018)Rapin and Teytaud]{nevergrad2018}
Rapin, J. and Teytaud, O.
\newblock {Nevergrad - A gradient-free optimization platform}.
\newblock \url{https://GitHub.com/FacebookResearch/Nevergrad}, 2018.

\bibitem[Shahriari et~al.(2016)Shahriari, Swersky, Wang, Adams, and
  de~Freitas]{boreview}
Shahriari, B., Swersky, K., Wang, Z., Adams, R.~P., and de~Freitas, N.
\newblock Taking the human out of the loop: A review of bayesian optimization.
\newblock \emph{Proceedings of the IEEE}, 104\penalty0 (1):\penalty0 148--175,
  2016.

\bibitem[Wang et~al.(2020)Wang, Fonseca, and Tian]{lamcts2020}
Wang, L., Fonseca, R., and Tian, Y.
\newblock Learning search space partition for black-box optimization using
  monte carlo tree search.
\newblock \emph{NeurIPS}, 2020.

\bibitem[Yang \& Shami(2020)Yang and Shami]{YANG2020295}
Yang, L. and Shami, A.
\newblock On hyperparameter optimization of machine learning algorithms: Theory
  and practice.
\newblock \emph{Neurocomputing}, 415:\penalty0 295--316, 2020.

\end{thebibliography}
\bibliographystyle{icml2022}

\end{document}